\documentclass[12pt]{article}

\usepackage[utf8]{inputenc}
\usepackage[T1]{fontenc}
\usepackage[margin=1in]{geometry}
\usepackage{setspace}
\usepackage{pdflscape}
\usepackage{amsmath, amssymb, amsfonts, mathtools, bbm}
\usepackage{graphicx, booktabs, multirow, float}
\usepackage{adjustbox, threeparttable, array, tabularx, dcolumn}
\usepackage{caption, subfig}
\usepackage{siunitx}
\usepackage{csquotes}
\usepackage[authoryear]{natbib}
\usepackage{hyperref}
\usepackage{cleveref}
\usepackage{amsthm}
\usepackage{appendix}

\newcolumntype{d}[1]{D{.}{.}{#1}}
\newcolumntype{Y}{>{\centering\arraybackslash}X}
\title{Let the Tree Decide: FABART \\ A Non-Parametric Factor Model}		
\author{Sofia Velasco\thanks{
European Central Bank and Queen Mary University London. I thank Haroon Mumtaz for valuable advice and guidance and am grateful for helpful suggestions received from Florian Huber, Karin Klieber, and Massimiliano Marcellino. 
The views expressed in this paper are those of the author and do not necessarily reflect those of the European Central Bank or the European System of Central Banks (ESCB). 
Correspondence: \texttt{s.m.velasco@qmul.ac.uk}.
}}
\date{\today}
\begin{document}
\maketitle
\begin{abstract}
\thispagestyle{plain}
This article proposes a novel framework that integrates Bayesian Additive Regression Trees (BART) into a Factor-Augmented Vector Autoregressive (FAVAR) model to forecast macro-financial variables and examine asymmetries in the transmission of oil price shocks. By employing nonparametric techniques for dimension reduction, the model captures complex, nonlinear relationships between observables and latent factors that are often missed by linear approaches. A simulation experiment comparing FABART to linear alternatives and a Monte Carlo experiment demonstrate that the framework accurately recovers the relationship between latent factors and observables in the presence of nonlinearities, while remaining consistent under linear data-generating processes. The empirical application shows that FABART substantially improves forecast accuracy for industrial production relative to linear benchmarks, particularly during periods of heightened volatility and economic stress. In addition, the model reveals pronounced sign asymmetries in the transmission of oil supply news shocks to the U.S. economy, with positive shocks generating stronger and more persistent contractions in real activity and inflation than the expansions triggered by negative shocks. A similar pattern emerges at the U.S. federal state level, where negative shocks lead to modest declines in employment compared to the substantially larger contractions observed after positive shocks.

\end{abstract}

\onehalfspacing
\setcounter{footnote}{0}
\renewcommand{\thefootnote}{\arabic{footnote}}

\section{Introduction}
Recent episodes of heightened volatility—driven by geopolitical uncertainty and the COVID-19 pandemic—have underscored the importance of capturing nonlinearities in economic time series models. In such periods, standard linear methods often fail to deliver reliable forecasts or accurate assessments of macroeconomic dynamics. Principal Component Analysis (PCA) and traditional linear factor models, widely used for dimensionality reduction in large datasets, impose restrictive linearity assumptions on the relationship between observed variables and latent factors. These limitations can weaken the predictive performance of factor-augmented frameworks, particularly in the presence of threshold effects, regime shifts, or extreme realizations.

To address these shortcomings, a growing literature has developed nonlinear dimension reduction techniques that are better equipped to uncover complex structures and enhance predictive accuracy. Early contributions by \cite{BaiNg2008} introduced quadratic principal components and targeted predictors, showing that even simple forms of nonlinearity can yield substantial forecasting gains. More recently, \cite{PelgerXiong2022} and \cite{hauzenberger2023real} highlight that allowing for state-dependent or nonlinear mappings between observables and latent factors can significantly improve model performance, particularly during recessionary episodes and periods of economic stress.

Building on these insights, this article proposes a novel Factor-Augmented Vector Autoregressive (FAVAR) model that integrates Bayesian Additive Regression Trees (BART) into the factor extraction stage—the FABART model. By introducing a nonparametric mapping between observed variables and latent factors, FABART captures complex, nonlinear relationships that linear methods may miss, while maintaining the interpretability of factor-augmented frameworks.

A simulation experiment that compares the model performance with a linear FAVAR model and Monte Carlo simulations demonstrate that FABART accurately recovers the relationship between latent factors and observables in the presence of nonlinearities without imposing spurious nonlinearities when the true data-generating process is linear. This robustness highlights its adaptability to different underlying structures and its ability to avoid overfitting.

The empirical application first assesses the forecasting performance of the FABART model relative to established linear benchmarks, with particular attention to periods of heightened volatility, such as the COVID-19 pandemic. Full-sample results show that FABART delivers forecast accuracy that is comparable to linear alternatives for real oil prices, but substantially improves forecasts for industrial production, particularly during periods of economic stress. For financial variables, forecast performance is more mixed: FABART delivers better density forecasts for the excess bond premium during periods of heightened volatility, but performs comparably to the linear BVAR in the case of the S\&P 500 index. Overall, the results suggest that forecasting performance deteriorated for both linear benchmarks—the BVAR and the linear FAVAR model—during and after the COVID-19 shock, whereas FABART remained broadly stable. These findings are consistent with recent evidence highlighting the advantages of flexible nonlinear models in capturing instability during periods of hightened turbulence (e.g., \cite{huber2020nowcasting}, \cite{hauzenberger2023real}, \cite{baumeister2024risky}), and underscore the relative strength of FABART in delivering more stable and reliable forecasts under these conditions.

In addition, this article investigates the transmission of oil supply news shocks by exploiting the flexible structure of FABART combined with Generalized Impulse Response Functions (GIRFs), which allow for a nonparametric assessment of nonlinearities. The results reveal pronounced sign asymmetries at the aggregate level: positive oil supply shocks induce stronger and more persistent contractions in real activity and inflation than the expansions triggered by negative shocks, consistent with the evidence documented by \cite{hamilton2003oil} and \cite{balke2002oil}. A similar pattern emerges for employment at the U.S. state level, with negative shocks generating weaker improvements in employment compared to the sharper contractions observed after positive shocks.\footnote{Moreover, the cross-sectional analysis highlights substantial heterogeneity in the magnitude of labor market responses: states with larger manufacturing sectors experience deeper employment contractions, while mining-intensive states are relatively less affected. These findings align with recent studies emphasizing the role of sectoral composition in amplifying the effects of macroeconomic shocks across regions (e.g., \cite{mumtaz2018state}).} 

The remainder of this article is structured as follows: Section \ref{sec:Econmeth} introduces the empirical model, estimation approach, and identification strategy. Section \ref{sec:Data} presents the Monte Carlo results. Section \ref{sec:Results} discusses the empirical findings. Section \ref{sec:Conclusion} concludes.

\section{Econometric methodology}\label{sec:Econmeth}
The factor-augmented vector autoregressive (FAVAR) model, introduced by \cite{Bernanke2005}, offers a reduced-dimensional vector autoregression (VAR) framework that integrates information from a large set of variables, addressing omitted-variable concerns by extracting latent factors from high-dimensional datasets. This method allows latent factors to co-evolve with observed variables. Extending this approach, I incorporate Bayesian additive regression trees (BART) to flexibly sample latent factors within a nonparametric framework, capturing complex, nonlinear dependencies while preserving interpretability. In line with other structural factor model frameworks, the Factor Bayesian Additive Tree (FABART) model leverages rank reduction techniques to expand the information set, facilitating the identification of structural disturbances without compromising degrees of freedom (\cite{korobilis2013assessing}).


\subsection{Model formulation}
In the following, the empirical framework is introduced through the factor-augmented VAR model by \cite{Bernanke2005} and extended to sample the factor loadings via Bayesian additive regression additive trees.

In the absence of non-parametric factor loadings the linear  FAVAR set-up is determined by the following measurement equation 
\begin{equation}\label{measurement}
	X_{i,t}= \Gamma_i Y_t + \epsilon_{i,t}
\end{equation}
where $X_{i,t}$ with i=$1,...,N$ denotes a large panel of N observable variables and $Y_t = \{Z_t,F^1_t, \dots, F^j_t\}$ is a matrix that contains $J=1,...,j$ latent factors that capture the shared movement patterns of the underlying series at each point in time. $Z_t$ is treated as an observed factor i.e. with factor loading equal to unity. In the application outlined in this paper the observed variable $Z_t$ is real oil prices.\footnote{In prominent frameworks that identify monetary policy shocks using FAVAR models (e.g., \cite{Bernanke2005}; \cite{baumeister2013changes}), the Federal Funds rate is typically treated as an observed factor. This assumption reflects its informational availability to both the econometrician and the monetary authority.} The relationship between the large granular dataset $X_{i,t}$ and the latent factors in $F_t$ is established by the factor loadings $\Gamma_i$ and $\epsilon_{i,t}$ are the i.i.d. errors with $E(\epsilon_{i,t}'\epsilon_{i,t})=\Sigma$, where $\Sigma$ is a diagonal matrix. 

The transition equation (\ref{transition}) with $M$ endogenous variables contained in matrix $Y_t$ of size $MxT$ and lag length $L=12$ is given by 
\begin{equation}\label{transition}
	Y_t = \sum_{l=1}^{L} \phi_{l,t} Y_{t-l} +\eta_t, \hspace{2mm} \eta_t \sim N(0,\Sigma)
\end{equation}

The objective of the FABART method is to examine the non-linearities in the estimation of the unobserved factors, specifically in the measurement equation (\ref{measurement}). In line with \cite{dolado2020quantile}, \cite{hauzenberger2023real}, and \cite{korobilis2024monitoring}, I aim to explore the non-parametric relationship between the factor and its loadings. Introducing such a flexible structure in the measurement equation enables mapping the high-dimensional dataset $X_{i,t}$ without imposing a restrictive linear relationship between the factors and the observables. This flexibility is essential for effectively reducing the dimensionality of large data panels while capturing complex dependencies in the data.

Several approaches have been proposed to introduce non-linearities in the relationship between latent factors and observables in $X_{i,t}$. Some methods achieve this by allowing for time-varying factor loadings (\cite{del2008dynamic}), while others introduce time variation in the factor dynamics (e.g., \cite{mumtaz2012evolving}; \cite{korobilis2013assessing}). My approach builds on this literature by explicitly modeling the non-parametric link between the factor and its loadings, offering a more flexible and data-driven framework for capturing structural changes in the underlying data relationships. Specifically, I postulate that $X_{t}=(X_{1}',..X_{N}')$ evolves according to a general multivariate model of the form 
\begin{equation}\label{measurement_nonpara}
	X_{t}= F(Y_t)+\eta_t, \hspace{2mm} \eta_{t} \sim N(0,\Sigma)
\end{equation}

where  $F(Y_t)= (f_1(Y_t),... f_N(Y_t)')$ is a N-dimensional vector of non-linear functions estimated non parametrically via Bayesian additive regression trees (BART).  

Each $f_i(Y_t)$ is approximated as   
\begin{equation} \label{fitted}
	f_i(Y_t) \approx \sum_{s=1}^{S}g_{is}(Y_t| \tau_{is},\mu_{iS})
 \end{equation}
with $g_{is}(Y_t| \tau_{is},\mu_{iS})$ being a single regression function, $f_i$ denoting the sum of the regression trees and $\tau_{is}$ are the tree structures related to the $i$th element in $X_{t}$. $\mu_{iS}$ are tree specific terminal nodes and $S$ measures the total number of trees.

As illustrated in \cite{huber2020nowcasting}, the space of each explanatory variable is recursively partitioned by every regression tree using binary splitting rules. Each rule involves selecting one explanatory variable and a threshold \( c \), determining whether an observation lies above or below that threshold. Based on these rules, observations are allocated to terminal nodes, which capture the fitted values conditional on the associated partition. These splitting rules can be written as \( \{ Y \in A_r \} \) if \( Y \) lies within the partition set \( A_r \) for \( r = 1, \dots, b \), and \( \{ Y \notin A_r \} \) otherwise.

Let \( \mathbf{Y}_{\cdot m} \) denote the \( m \)-th column of \( Y \), containing the latent factors. A binary split is defined as:
\begin{equation}\label{Splitting1}
\begin{split}
 \mathbf{Y}_{\cdot m} \leq c , \\
\mathbf{Y}_{\cdot m} > c
\end{split}
\end{equation}

For each of the \( N \) variables in \( X\), the fitted values from a regression tree are obtained via the following step function:
\begin{equation}\label{step}
g(Y \mid \tau, \mu) = \sum_{r=1}^{b} \mu_{r} \, \mathbbm{1}{ (Y \in A_r) },
\end{equation}

where \( \mathbbm{1} \{ \cdot \} \) denotes the indicator function, which takes the value 1 if \( Y \) falls within the partition region \( A_r \) defined by the sequence of splitting rules encoded in the tree structure. Each component of the regression tree is treated as an unknown parameter and estimated from the data. The parameter set includes: the terminal node values \( \mu \), the number of terminal nodes \( b \), and all elements defining the tree structure—namely, the splitting variable \( \mathbf{Y}_{\cdot m} \) and the threshold \( c \) that governs the binary split.

\subsection{Estimation}
\subsubsection{The Priors}
The prior distributions introduced by \cite{chipman2010bart} are fundamental in mitigating the risk of overfitting. The joint prior for the parameters governing the \(S \) trees at horizon \( h \) is decomposed as:

\begin{equation}
    p((T_1^{(h)}, \mu_1^{(h)}), (T_2^{(h)}, \mu_2^{(h)}), \dots, (T_s^{(h)}, \mu_s^{(h)})) = \prod_{s=1}^{s} p(\mu_s^{(h)} | T_s^{(h)}) p(T_s^{(h)}),
\end{equation}

where the conditional prior follows:

\begin{equation}
    p(\mu_s^{(h)} | T_s^{(h)}) = \prod_{b=1}^{B_s} p(\mu_{b_s}^{(h)} | T_s^{(h)}).
\end{equation}

The structure of the tree \( T_s^{(h)} \) depends on the likelihood that a given node at depth \( d = 0,1,2, \dots \) is an internal node rather than a terminal one. This probability is determined as $\alpha(1 + d)^{-\beta}$, with \( \alpha \in (0,1) \) and \( \beta > 0 \). Higher values of \( \beta \) and lower values of \( \alpha \) strengthen the prior belief that the tree structure remains simple (i.e., shorter). As recommended by \cite{chipman2010bart}, I set \( \alpha = 0.95 \) and \( \beta = 2 \). The threshold parameter \( c \) follows a uniform distribution over the observed range of the variables. By default, the selection of splitting variables is assumed to be uniform across regressors.

To define \( p(\mu_s^{(h)} | T_s^{(h)}) \), \cite{chipman2010bart} propose rescaling the dependent variable such that its values fall within the interval \([-0.5, 0.5]\). Consequently, the function \( m_h(x_t, z_t, w_{t+h}^{(h)}) \) is expected to remain within this range. The prior distribution of \( p(\mu_s^{(h)} | T_s^{(h)}) \) follows a normal distribution \( N(0,S) \), where the variance \( S \) is defined as $ S = \frac{1}{2\kappa(0.5)^2}$, as recommended by \cite{chipman2010bart} \( \kappa = 2 \) . Given this prior, the probability that the conditional mean of the dependent variable lies between \([-0.5,0.5]\) is approximately 95\%.

For the variance term \( \sigma_{j} \), I adopt a conjugate inverse \( \chi^2 \) prior. The hyperparameters of this prior are specified using an estimated variance \( \hat{\sigma}_{j} \) derived from a linear regression. As noted in \cite{mumtaz2022impulse}, when the model exhibits non-linearity, this estimate is typically biased upwards. 
In line with common practice, the hyperparameters are chosen to satisfy \( P(\sigma_j < \hat{\sigma}_j) = 0.75 \) and $\nu_j = T/2$ . The number of trees \( S \) is fixed and set to 250, e.g., see \cite{chipman2010bart} or \cite{huber2020nowcasting}).
\begin{table}[h]
    \centering
    \begin{threeparttable}
        \caption{Summary of prior hyperparameters.}
        \label{tab:prior_hyperparameters}
        \begin{tabular}{lll}
            \toprule
            \textbf{Description} & \textbf{Parameters} & \textbf{Hyperparameters} \\
            \midrule
            Trees  & $\alpha(1 + n)^{-\beta}$ & $\alpha = 0.95$ \\  
            \textit{(probability of non-terminal node)}& & $\beta = 2$ \\  
            Terminal nodes & $\mu_{i,j}|\mathcal{T}_{ij} \sim \mathcal{N}(0, \sigma_{\mu_j}^2)$ & $\sigma_{\mu_j} = \frac{(\max(\mathbf{X}_{\cdot j}) - \min(\mathbf{X}_{\cdot j}))}{(2\sqrt{\nu S})}$ \\  
            & & $\gamma = 2$ \\  
            & & $S = 250$ (number of trees) \\  
            Error variances & $\sigma_j^2 \sim \nu_j \xi_j / \chi^2_{\nu_j}$ & $\nu_j = T/2$ \\  
            & $\xi_j \text{ s.t. } P(\sigma_j < \hat{\sigma}_j) = v$ & $v = 0.75$ \\  
            Covariances & $q_{jl} | \tau_{jl}, \lambda \sim \mathcal{N}(0, \tau_{jl}^2 \lambda^2)$ & $\tau_{jl} \sim \mathcal{C}^+(0,1)$ \\  
            & & $\lambda \sim \mathcal{C}^+(0,1)$ \\  
            \bottomrule
        \end{tabular}

        \begin{tablenotes}
        \tiny{
            \item \textit{Notes:} Description provides information about the respective model parameter, while Parameters define the corresponding probability distributions. The Hyperparameters column specifies the chosen values for these distributions. In this notation, $X_{\cdot j}$ refers to the $j$th column of $X$. The initial variance, $\hat{\sigma}^2$, is estimated using a linear regression model. The same prior specification is maintained across the simulation study and empirical application to ensure consistency in estimation.}
        \end{tablenotes}
    \end{threeparttable}
\end{table}

\subsubsection{Posterior simulation}
The Bayesian "sum-of-trees" framework developed by \cite{chipman2010bart} enables posterior sampling through an iterative Bayesian backfitting MCMC algorithm. This approach regularizes each tree using a prior, ensuring that individual trees act as weak learners.\footnote{Although conceptually similar, BART differs technically from the gradient boosting method of \cite{friedman2001greedy} by incorporating a prior to constrain the trees and applying Bayesian backfitting across a fixed number of trees.} 

The sampling of each tree structure \( T_{js} \), with index \( j = 1, \dots, N \) denoting the different equations of the model, follows the Bayesian backfitting algorithm introduced by \cite{chipman2010bart}. The posterior distribution of the trees is computed by integrating out the terminal node parameters \( \mu_{js} \), and is given by:

\begin{equation}
p(\tau_{js} \mid R_{js}, \sigma_j) \propto p(\tau_{js}) 
\int p(R_{js} \mid \mu_{js}, \tau_{js}, \sigma_j) p(\mu_{js} \mid \tau_{js}, \sigma_j) d\mu_{js}.
\end{equation}

Here, the partial residual vector \( R_{js} \) is defined as:

\begin{equation}
R_{js} = X_{\cdot j} - \sum_{s \neq s} g_{js}(Y \mid \tau_{js}, \mu_{js}) 
\end{equation}

The likelihood term \( p(R_{js} \mid \tau_{js}, \sigma_j) \) facilitates efficient estimation by marginalizing over \( \mu_{js} \).

Tree structures are sampled using the Metropolis-Hastings (MH) algorithm, as outlined in \cite{chipman1998bayesian}. A candidate tree \( \tau_{js}^* \) is proposed from a distribution \( q(\tau_{js}^i, \tau_{js}^*) \) based on the current tree state \( \tau_{js}^i \). The acceptance probability for \( \tau_{js}^{i+1} \) is given by:

\begin{equation}\label{acceptanceprob}
a(\tau_{js}^i, \tau_{js}^*) = \frac{q(\tau_{js}^*, \tau_{js}^i) p(R_{js} \mid \tau_{js}^*, \sigma_j) p(\tau_{js}^*)}
{q(\tau_{js}^i, \tau_{js}^*) p(R_{js} \mid \tau_{js}^i, \sigma_j) p(\tau_{js}^i)}.
\end{equation}
The proposal distribution \( q(\tau_{js}^i, \tau_{js}^*) \) incorporates four types of moves:
\begin{enumerate}
    \item \textit{Grow}: Expands a terminal node into two new child nodes (probability 0.25).
    \item \textit{Prune}: Merges two terminal nodes back into a single parent node (probability 0.25).
    \item \textit{Change}: Modifies the splitting rule at an internal node (probability 0.40).
    \item \textit{Swap}: Exchanges decision rules between a parent and child node (probability 0.10).
\end{enumerate}

By integrating out \( \mu_{js} \), this approach maintains computational efficiency while ensuring a stable parameter space for posterior inference.

To fully implement the MCMC algorithm the  latent states conditional distribution of the BART parameters need to be drawn. This is akin to the strategy employed by \cite{huber2022inference}, who extend the Bayesian Additive Vector Autoregression Trees (BAVART) model by integrating latent state sampling to accommodate mixed-frequency data. Specifically, leveraging insights from the literature on effect size estimation in black-box models (see \cite{crawford2019variable}), a linear approximation to $F(Y_t)$ is constructed. This approximation enables the efficient estimation of latent states using conditionally Gaussian methods.

This linear approximation was initially introduced in the context of Gaussian kernel (GK) and Gaussian process (GP) regressions. However, its applicability extends far beyond these methods and is relevant to a broad spectrum of machine learning and econometric techniques, see \cite{ish2019interpreting}. The primary requirement is the estimation of a nonlinear function $F = (F(Y_1), \dots, F(Y_T))$ over $T$ observations.

To illustrate the concept of linear approximation, consider a standard linear regression model where the effect size (or regression coefficient) quantifies the extent to which $Y$, containing both latent factors and observables, projects onto the granular dataset $X_{i}$. 

This projection is defined as:
\(
\hat{A_i} = \text{Proj}(Y, X_{i}) = Y^{\dagger} X_{i}\), where $Y^\dagger$, a $K \times T$ matrix, represents the Moore-Penrose inverse. If $Y$ is of full rank, the projection simplifies to $(Y'Y)^{-1}Y'X_{i}$, making the effect size equivalent to the least squares estimate, which captures the marginal impact of explanatory variables on dependent variables.

Building on \cite{crawford2018bayesian} and \cite{crawford2019variable}, this approach is extended by projecting $X_i$ onto the matrix $F$, yielding the following effect size estimate:

\( \tilde{A_i} = \text{Proj}(F, X_i) = F^\dagger X_i\).

The core intuition behind this approximation is that regressing $X_i$ onto $F$ provides insight into how much of its variance is accounted for by $F$. This offers a clear interpretation of the relationship between each factor $F_i, (i = 1, \dots, K)$ and the corresponding $X_i$. Without additional regularization, the projection naturally implies that $F\tilde{A_i} \approx X_i$.

 $\tilde{A_i}$ allows to produce a linear approximation to the non-parametric multivariate model and obtain the elements of $\Gamma$ Factor loadings matrix for each $N$-variable from the measurement equation (\ref{measurement}), such that 
\begin{equation}
	X_{i,t}= \tilde{A_i}' Y_t + \epsilon_{i,t}
\end{equation}
In matrix notation this would correspond to 
\begin{equation}
\begin{pmatrix}
Z_t\\
X_{1,t} \\
\vdots \\
X_{N,t}
\end{pmatrix}
=
\begin{pmatrix}
1 & 0 & \dots & 0 & 0\\
\Psi^{11} & \Lambda^{11} & \Lambda^{12} & \dots & \Lambda^{1j}\\
 \Psi^{21} &\Lambda^{21} & \Lambda^{22} & \dots & \Lambda^{2j} \\
\vdots & \vdots & \ddots & \vdots & \vdots \\
 \Psi^{N1}&\Lambda^{N1} & \Lambda^{N2} & \dots & \Lambda^{Nj}

\end{pmatrix}
\begin{pmatrix}
Z_t\\
F^1_t \\
F^2_t \\
\vdots \\
F^j_t 
\end{pmatrix}
+
\begin{pmatrix}
0\\
e_{1,t} \\
e_{2,t} \\
\vdots \\
e_{N,t}

\end{pmatrix},
\end{equation}

where $\Lambda$ and $\Psi$ are proxied by $\tilde{A_i}$, representing the elements of the factor loading matrix $\Gamma$, which has dimensions $(N+1) \times (J+1)$. The structure of this matrix allows certain variables to exhibit a contemporaneous relationship with the observable variable $Z_t$, as indicated by the non-zero elements of $\Psi$.


Based on this approximation to a linear model with Gaussian shocks the latent factors are drawn based on the \cite{carter1994gibbs} algorithm as in \cite{mumtaz2010evolving}.

The algorithm for sampling from the full conditional posterior distribution is detailed below. The MCMC algorithm sequentially draws samples from the full conditional posterior distributions described in the preceding subsections. Given a set of initial values, the algorithm follows these steps in each iteration\footnote{Steps 2., 3. and \textit{6.} are implemented using the R package dbarts.}:

\begin{enumerate}
\item Conditional on a draw of the unobserved components sample VAR coefficients from a normal posterior distribution.\footnote{ A natural conjugate prior for the VAR parameters is introduced as in \cite{banbura2010large}. For more details on the prior settings refer to the Appendix.} 
\item Sample the $S$ trees, denoted as $\mathcal{T}_{i}^{s}$, for each equation using the Metropolis–Hastings algorithm. The acceptance probability is determined by equation (\ref{acceptanceprob}), and the proposed moves follow the distribution $q(\mathcal{T}_{i}^{s}, \mathcal{T}_{i}^{s*})$. 
    \item Draw the terminal node parameters linked to each tree and equation $i$ from a Gaussian distribution.
  
   \item Obtain the factor loadings through a linear approximation to $F(Y_t)$ (\cite{crawford2019variable}; \cite{huber2020nowcasting}).
   \item Conditional on the draws obtained in Steps 1.-5. apply the \cite{carter1994gibbs} algorithm to cast the factors in a state space model.
   \item \textit{If the FABART model is used to forecast the path of $X_t$ for the $NN$ variables in the sample, the predictions will be conditional on the forecast of $Y_t$ over the selected horizon.}
\end{enumerate}

The algorithm generates a total of 30,000 draws, with the initial 15,000 discarded as burn-in. 

\subsection{Identification }
\subsubsection{Generalized impulse response functions}

Given the nonlinear nature of the model, impulse responses are obtained using the Generalized Impulse Response Function (GIRF) framework proposed by \cite{koop1996impulse}.
The GIRF at horizon $k$ is defined as:

\begin{equation}
\text{GIRF}_{ij}(k, \Psi_t, e_j,\overline{Y}) = E[Y_{i,t+k} | \Psi_t, \overline{Y}, e_j] - E[Y_{i,t+k} | \Psi_t, \overline{Y}],
\end{equation}

where $\Psi_t$ denotes the set of model parameters, and $e_j$ represents the structural shock. The long-run mean of the data, denoted as $\overline{Y}$, is given by:

\begin{equation}
\overline{Y} = \left( I - \bar{F} \right)^{-1} \hat{C}
\end{equation}

By using the long-run mean instead of the historical values of the series, the influence of specific data point selections should be reduced. Since the model is nonlinear, the impulse responses are computed via Monte Carlo simulation, accounting for history-dependent effects and potential asymmetries in shock transmission. 

\subsubsection{Identification of Oil price shocks OLD}
The covariance structure of the reduced-form residuals, denoted as $\boldsymbol{\Sigma}$, is specified as:

\begin{equation}
\boldsymbol{\Sigma} = \mathbf{A} \mathbf{Q} \mathbf{A}',
\end{equation}

where $\mathbf{A}$ represents a lower triangular matrix and $\mathbf{Q}$ is an orthogonal matrix satisfying $\mathbf{Q}'\mathbf{Q} = \mathbf{I}$. The structural shocks, $\boldsymbol{\varepsilon}_t$, are then obtained as:

\begin{equation}
\boldsymbol{\varepsilon}_t = \mathbf{A}^{-1} \mathbf{u}_t
\end{equation}

where $\mathbf{u}_t$ represents the reduced-form errors. Following \cite{kanzig2021macroeconomic}, the identification of an oil supply news shock is established by assuming it corresponds to the first element of $\boldsymbol{\varepsilon}_t$, i.e., $\varepsilon_{1t}$. This is achieved through the use of an instrumental variable, $m_t$, satisfying:

\begin{equation}
m_t = \beta \varepsilon_{1t} + \sigma v_t, \quad v_t \sim \mathcal{N}(0,1) \label{Instrument}
\end{equation}

with the orthogonality condition $E(m_t \varepsilon_{jt}) = 0$ for all $j \neq 1$. 
\begin{equation}
\rho^2 = \frac{\beta^2}{\beta^2 + \sigma^2}.
\end{equation}
In the empirical application, the instrumental variable $m_t$ follows \cite{kanzig2021macroeconomic} and is constructed based on variations in oil futures prices around OPEC announcements.

\subsubsection{Identification of Oil Price Shocks}

The covariance structure of the reduced-form residuals, denoted as $\boldsymbol{\Sigma}$, is specified as:

\begin{equation}
\boldsymbol{\Sigma} = \mathbf{A} \mathbf{Q} \mathbf{A}',
\end{equation}

where $\mathbf{A}$ represents a lower triangular matrix and $\mathbf{Q}$ is an orthogonal matrix satisfying $\mathbf{Q}'\mathbf{Q} = \mathbf{I}$. The structural shocks, $\boldsymbol{\varepsilon}_t$, are then obtained as:

\begin{equation}
\boldsymbol{\varepsilon}_t = \mathbf{A}^{-1} \mathbf{u}_t,
\end{equation}

where $\mathbf{u}_t$ represents the reduced-form errors. Following \cite{kanzig2021macroeconomic}, the identification of an oil supply news shock is established by assuming it corresponds to the first element of $\boldsymbol{\varepsilon}_t$, i.e., $\varepsilon_{1t}$. This is achieved through the use of an instrumental variable, $m_t$, satisfying:

\begin{equation}
m_t = \beta \varepsilon_{1t} + \sigma v_t, \quad v_t \sim \mathcal{N}(0,1), \label{Instrument}
\end{equation}

with the orthogonality condition $E(m_t \varepsilon_{jt}) = 0$ for all $j \neq 1$. The strength of the instrument is summarized by the reliability ratio:
\begin{equation}
\rho^2 = \frac{\beta^2}{\beta^2 + \sigma^2}.
\end{equation}

This is carried out using a frequentist approach. Importantly, equation~\eqref{Instrument} is not introduced as an additional equation in the model. Instead, for each draw of the transition equation parameters, the first column of $\mathbf{A}$ is computed by projecting $m_t$ on the reduced-form residuals using the standard frequentist method, as in \cite{miranda2023identification}.

In the empirical application, the instrumental variable $m_t$ follows \cite{kanzig2021macroeconomic} and is constructed based on variations in oil futures prices around OPEC announcements.

\section{Simulation experiment}\label{sec:Data}
The primary objective of this section is to assess the FABART model’s ability to capture nonlinear dynamics in economic data. To this end, I develop a simulation experiment that generates data reflecting nonlinear features commonly observed in macroeconomic time series. The central question is whether FABART can successfully recover these nonlinear relationships without imposing restrictive parametric assumptions.

Model performance is evaluated within a factor model framework by comparing outcomes across three data-generating processes (DGPs): one linear and two nonlinear.

The factor evolves according to an autoregressive process of order three:
\begin{equation}
    F_t = c + b_1 F_{t-1}+ b_2 F_{t-2}+ b_3 F_{t-3} + e_t, \quad e_t \sim \mathcal{N}(0, \sigma^2)
\end{equation}
where \( e_t \) is normally distributed with variance \(\sigma^2 \). The coefficient vector \( \beta \) is parameterized as \( \beta = (0.6, -0.3, 0.2) \). The setup follows a standard factor model in which the $N$=20 observed variables \( X_t \) is driven by $J$=1 latent factor \( F^{J}_t \). The baseline (linear) specification is given by:
\begin{equation}\label{MCeq11}
    X_t = B F^{J}_t + V_t, \quad V_t \sim \mathcal{N}(0, R),
\end{equation}
where  \( X_t \) is a \( 20 \times T \) vector of observed variables, \( B \) is a \(20 \times 1 \) factor loading matrix, and \( V_t \) follows a normal distribution with mean zero and diagonal covariance matrix \( R \) (\( 20 \times 20 \)). The elements of \(B\) are drawn independently from a uniform distribution over the interval \([-0.9, 0.9]\). The diagonal elements of \(R\), representing the idiosyncratic variances of each series, are independently sampled from a uniform distribution.

To introduce nonlinearities, Equation \ref{MCeq11} is extended in two ways:  

\begin{itemize}  
    \item[(i)] \textbf{Quadratic nonlinearity in the factor loadings:}  
    The factor loading is squared to capture asymmetric amplification effects:  
    \begin{equation}\label{nonlinear1}
        X_t = B^2 F_t + V_t.  
    \end{equation}  

    \item[(ii)] \textbf{Hyperbolic tangent transformation for bounded nonlinear effects:}  
    A hyperbolic tangent transformation is applied to the factor loading to allow for smooth saturation at extreme values, mimicking nonlinear adjustment dynamics observed in financial and macroeconomic series:\footnote{The use of \(\tanh\) in time series modeling has been explored in nonlinear autoregressive models \cite{Mohammed2022} and machine learning-based forecasting \cite{Sami2021}. Additionally, \cite{BIS2023} demonstrate the effectiveness of neural network-based approaches, which frequently employ the \(\tanh\) function, in estimating nonlinear heterogeneous agent models.}  
    \begin{equation}\label{nonlinear2}
        X_t = \tanh(B F_t) + V_t.  
    \end{equation}  
\end{itemize}

  To evaluate model performance in recovering the latent factor, a recursive one-step-ahead forecasting exercise is conducted across all synthetic datasets. The factor process $F_t$ is held constant across data-generating processes to isolate differences in model behavior, allowing for a direct comparison between FABART, a linear FAVAR, and a Random Walk (RW) benchmark. Table~\ref{tab:RMSEMC} reports the Root Mean Squared Errors (RMSEs) for each method, normalized relative to the RW benchmark, which is scaled to 1. Across all specifications, the FABART model delivers the lowest forecast errors.
  
\begin{table}[htp]
\caption{RMSE of unobserved factor}
\label{tab:RMSEMC}
\centering
\begin{threeparttable}
\begin{tabular}{lccc}
    \toprule
    \textbf{Data-generating process} & \textbf{FABART} & \textbf{FAVAR} & \textbf{RW} \\
    \midrule
    Linear       & 0.641 & 0.652 & 1.000 \\
    Nonlinear I  & 0.648 & 0.655 & 1.000 \\
    Nonlinear II & 0.650 & 0.655 & 1.000 \\
    \bottomrule
\end{tabular}
\begin{tablenotes}
    \tiny
    \item \textit{Notes:} RMSEs for $F_t$ under FABART and FAVAR are normalized by the Random Walk (RW) benchmark (RMSE = 1.334), which is scaled to 1. Lower values indicate better performance relative to RW.
\end{tablenotes}
\end{threeparttable}
\end{table}

\begin{landscape}
\begin{figure}[htp]
    \centering
    \begin{minipage}[b]{0.46\linewidth}
        \centering
        \includegraphics[width=\textwidth]{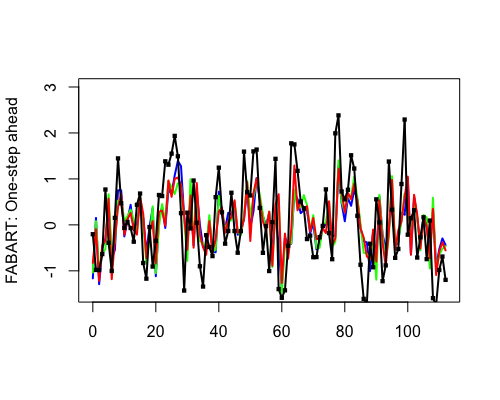}
        \includegraphics[width=\textwidth]{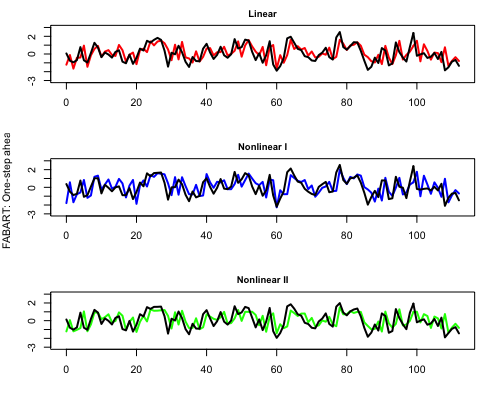}
    \end{minipage}
    \hfill
    \begin{minipage}[b]{0.46\linewidth}
        \centering
        \includegraphics[width=\textwidth]{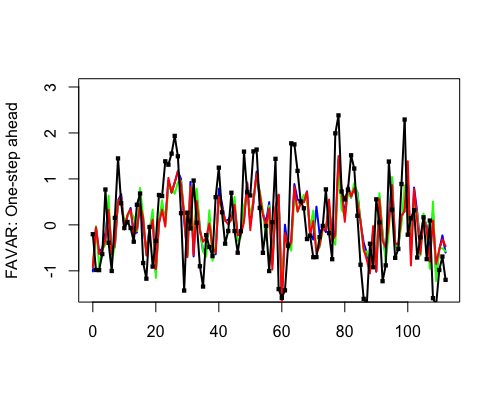}
        \includegraphics[width=\textwidth]{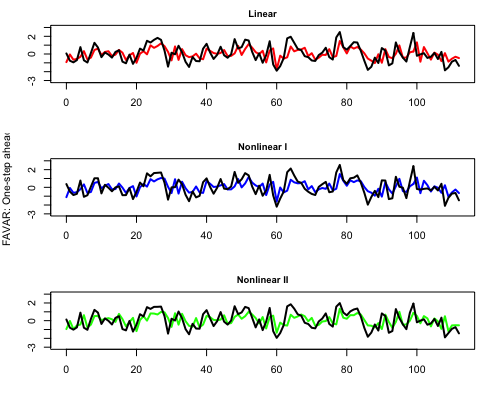}
    \end{minipage}
    
    \tiny{
    \textit{Note:} One-step-ahead projections obtained using the FABART and FAVAR methodologies. Black lines indicate the true data-generating process (DGP), straight lines the median estimates. Red lines correspond to projections under the linear model; blue lines reflect results based on the DGP with quadratic nonlinearities; and green lines correspond to the DGP incorporating a hyperbolic tangent transformation. The top panels show projections for the latent factor $F_t$, while the bottom panels display projections for one of the observable variables $X_t$.}
    \caption{Single realization from the DGP. One-step-ahead projections under FABART (left) and FAVAR (right).}
    \label{fig:MC_combined_onestep}
\end{figure}
\end{landscape}

Figure~\ref{fig:MC_combined_onestep} displays the one-step-ahead forecast paths for the latent factor $F_t$ and a representative observable $X_t$, comparing predictions from the FABART and linear FAVAR models. Visually, the FABART forecasts align more closely with the true series, particularly under nonlinear data-generating processes, highlighting the model’s flexibility in capturing complex dynamics.

A Monte Carlo experiment based on 100 replications of each data-generating process (DGP)—in which the latent factor is held fixed while the observable variables \(X_t\) vary across replications—provides a deeper assessment of the properties of the FABART model. The results of this exercise are presented in the Annex. Figure \ref{fig:distributionFE} displays the distribution of estimation errors for the latent factor \(F_t\), obtained using the FABART methodology. The first column illustrates the posterior estimates of \(F_t\) across replications, while the second column shows the corresponding estimation errors, defined as the difference between each posterior draw and the true factor. The figure highlights that, on average, errors are centered around zero across all DGPs and time periods. Estimation is most accurate in the linear case, with slightly larger errors observed under the nonlinear DGPs. This is reflected in high average correlations between the estimated and true factors: 0.982 for the linear DGP, 0.968 for the quadratic nonlinear DGP, and 0.966 for the hyperbolic tangent case. These findings confirm that the proposed methodology accurately recovers the latent factor across varying data-generating environments.
\newpage

\section{Empirical Application}\label{sec:Results} 
The analysis in this section applies the FABART methodology to evaluate its forecasting performance across a range of macro-financial variables—including oil prices, industrial production, and financial market indicators—and to examine the transmission of oil price shocks to the U.S. economy. The results are presented in two parts.

The first part assesses the model’s forecasting accuracy relative to established benchmarks, with particular attention to periods of heightened volatility, such as the COVID-19 pandemic, to evaluate its ability to capture complex macroeconomic dynamics.

The second part investigates nonlinearities in the transmission of oil price shocks, providing empirical evidence on how the sign of such shocks affects key macroeconomic outcomes in the U.S. economy on aggregate and on the differences at the state level dimension.

\subsection{Dataset}
The baseline specification exploits the potential of rank-reduction techniques by incorporating a set of 187 macroeconomic and financial time series. This approach enables an in-depth analysis of oil price dynamics, leveraging variables emphasized in recent contributions to the literature—such as \cite{baumeister2015forecasting}, \cite{kanzig2021macroeconomic}, and \cite{baumeister2024risky}. It also expands the information set relevant to key determinants of oil price dynamics (Table \ref{tab:Oilmarketfund}) by including the full FRED-MD monthly database of macroeconomic indicators proposed by \cite{mccracken2016fred}. A detailed list of the FRED-MD variables and transformation codes is provided in Table \ref{fred} located in the Apppendix. To facilitate a state-level analysis of the employment response to oil price shocks, the dataset further includes total nonfarm employment for all U.S. states.

This broad information set facilitates a richer understanding of the structural and cyclical drivers of oil prices by capturing a wide range of economic and financial conditions. At the same time, the use of dimension reduction techniques ensures that the resulting framework remains computationally tractable, even in high-dimensional settings, while still allowing for the analysis of the dynamics and interactions of the broader set of included macroeconomic and financial series.

\subsubsection{Key determinants of oil price dynamics}

A key variable of interest in this study is the real price of WTI crude oil, a widely used global benchmark for crude and refined petroleum products. To obtain a consistent measure of real oil price dynamics, Brent prices are deflated using the U.S. Consumer Price Index (CPI).

The selection of predictor variables reflects their theoretical and empirical relevance in capturing the fundamental drivers of oil prices, including supply conditions, demand pressures, and broader macroeconomic influences. This setup is motivated by the extensive literature on oil price modeling, which emphasizes the interplay between structural and cyclical components.

For instance, \cite{baumeister2015forecasting} highlight the importance of including oil production, inventory levels, and indicators of economic activity to adequately capture both supply- and demand-side shocks. Their findings underscore the role of global economic momentum and inventory dynamics in shaping oil price fluctuations, particularly in capturing nonlinear and episodic responses to macroeconomic and geopolitical developments.

On the supply side, global oil production and petroleum inventories serve as key indicators of physical market fundamentals, reflecting adjustments in output and stockpiles that buffer short-run imbalances. On the demand side, global industrial production and composite indicators of economic activity capture fluctuations in energy consumption over the business cycle. Together, these variables provide a comprehensive basis for modeling the dynamics of real oil prices across regimes and time horizons.

Table \ref{tab:Oilmarketfund} provides an overview of the relevant oil market drivers considered in the analysis.

In selecting the dataset for this analysis, I follow \cite{baumeister2015forecasting}, \cite{kanzig2021macroeconomic}, \cite{miescu2024cardiff}, and \cite{baumeister2024risky}. The partial dependence analysis by \cite{baumeister2024risky} highlights key nonlinear relationships influencing oil price predictions. While moderate changes in predictors like global fuel consumption and petroleum inventories have limited effects, extreme variations—such as sharp contractions in the GECON indicator, steep declines in rig counts, or significant increases in the gasoline-Brent spread—induce disproportionate and asymmetric impacts on oil prices.

All variables are measured at a monthly frequency and are transformed into growth rates by taking the first difference of their natural logarithms. Exceptions to this transformation include the GECON indicator, which is inherently stationary due to its construction. The transformation of real oil prices differs across empirical applications: for the forecasting exercise, the series is included in levels, consistent with the approach in \cite{baumeister2024risky}, while the structural analysis employs the log difference of real oil prices, in line with the specification used by \cite{miescu2024cardiff}.

\begin{table}[htb!]
\centering
\caption{Oil Market Fundamentals (1974M1--2024M08)}
\label{tab:Oilmarketfund}
\begin{threeparttable}
\begin{tabularx}{\textwidth}{lXl}
\hline
\textbf{Variable} & \textbf{Description} & \textbf{Source} \\ 
\hline
Real Oil Price & WTI spot price, deflated by U.S. CPI. & EIA\tnote{a}, FRED\tnote{b} \\ 
Global Oil Production & Total global oil production. & IES\tnote{c} \\ 
OECD Petroleum Inventories & Crude oil and petroleum product inventories in OECD countries. & EIA\tnote{a}, IES\tnote{c} \\ 
Global Industrial Production & Monthly global industrial output. \small{\cite{baumeister2019structural}} &  webpage\tnote{d} \\ 
GECON Indicator & Global economic conditions indicator, reflecting economic, financial, and geopolitical factors. \small{\cite{baumeister2022energy}} & webpage\tnote{d} \\ 
\hline
\end{tabularx}
\begin{tablenotes}
\footnotesize
\item[a] Energy Information Administration (EIA).
\item[b] Federal Reserve Economic Data (FRED).
\item[c] International Energy Statistics (IES).
\item[d] Christiane Baumeister's webpage (\url{https://sites.google.com/site/cjsbaumeister/datasets}).
\end{tablenotes}
\end{threeparttable}
\end{table}

To benchmark the performance of the FABART model, a linear BVAR is estimated using a selected subset of macroeconomic and oil market variables. These include the real WTI oil price, world oil production, world oil inventories, global industrial production, U.S. industrial production, and the U.S. Consumer Price Index (CPI). A detailed description of these variables is provided in Table \ref{tab:redvariables} in the Annex. In addition, a linear FAVAR model is estimated using the same variable set as the FABART model, allowing for a direct comparison between the nonlinear and linear factor-augmented approaches.

\subsubsection{State level data }\label{statelevel}
To investigate cross-state heterogeneity in the transmission of oil price shocks, the analysis incorporates state-level nonfarm employment data. These series, sourced from the U.S. Bureau of Labor Statistics and accessed via the Federal Reserve Bank of St. Louis FRED database, provide a consistent measure of labor market conditions across all U.S. states. Each series is first seasonally adjusted using standard procedures based on the X-13ARIMA-SEATS methodology, and then transformed by computing the logarithm of the first difference to ensure stationarity and focus on cyclical dynamics. 

In addition, to assess how differences in economic structure influence the transmission of oil price shocks, the analysis follows \cite{mumtaz2018state} and draws on data describing the industrial composition of each state. Specifically, the dataset includes annual state-level GDP by industry from 1963 to 2013, averaged over time to mitigate short-run volatility and emphasize long-run structural characteristics. These data are provided by the U.S. Bureau of Economic Analysis (BEA), with industry classifications based on the Standard Industrial Classification (SIC) system through 1996 and the North American Industry Classification System (NAICS) from 1997 onward.

\subsection{Forecast Evaluation}
This subsection evaluates the forecasting performance of the FABART model relative to two linear benchmarks: a Bayesian Vector Autoregression (BVAR) estimated on a selected subset of macroeconomic and oil market variables, and a Factor-Augmented VAR (FAVAR) that uses the same variables and factor dimension as the FABART model. Forecast accuracy is assessed using both root mean squared errors (RMSEs) and predictive densities estimated via kernel methods.

Evaluation of predictive densities via kernel methods allows for the capture of non-normal features such as asymmetry and fat tails, which tend to become more pronounced at forecast horizons beyond one month (see \cite{geweke2010comparing}). This evaluation criterion is selected because, while the posterior predictive densities generated by linear models—such as Bayesian Vector Autoregressions (BVARs) and Factor-Augmented VARs (FAVARs)—are Gaussian by construction, the FABART model accommodates richer distributional characteristics. This distinction becomes especially relevant in periods of heightened uncertainty, where there is evidence of deviations from normality in economic time series (see \cite{alessandri2017financial}).

Formally, let $M = M_0$ denote the benchmark model, the FABART model introduced in this study. Two alternative models are considered for comparison, denoted as $M_1$ and $M_2$: (1) a fixed-coefficient Bayesian VAR, which assumes time-invariant relationships among variables; and (2) a linear FAVAR model, which incorporates latent factor structures while maintaining Gaussian density assumptions. For both the linear FAVAR and the FABART, seven factors are considered ($J = 7$),
which explains approximately 47\% of the total variance in the dataset. This choice is consistent with the findings of \cite{stock2005implications}, who estimate seven dynamic factors using a comparable macroeconomic dataset, and \cite{mccracken2016fred}, who report similar explanatory power in their FRED-MD-based factor model. For all three specifications, the sample spans 1974:M01–2024:06 with a lag length of $L = 12$. The final observation used for estimation is 2023:06, with forecast evaluation conducted over the period 2023:07–2024:06. As is standard for US data, the overall prior tightness is set to $\iota = 0.1$. As in \cite{banbura2010large}, the tightness of the sum of coefficients prior is set as $\lambda = 10\iota$, and a flat prior is imposed on the constant term (see, e.g., \cite{alessandri2017financial}).

For each model $M \in \{M_0, M_1, M_2\}$, the log predictive likelihood is computed as:
\begin{equation}
    \log p(Z_T | Z_S, M) = \sum_{t=S+1}^{T-h} \log p(Z_{t+h} | Z_t),
\end{equation}
where $\log p(Z_{t+h} | Z_t)$ denotes the predictive likelihood of a variable of interest, $h$ is the forecast horizon, and $t = S+1, \dots, T-h$ defines the forecast evaluation window, with $S < T$.

I evaluate forecast accuracy across one-, three-, and twelve-month-ahead horizons using root mean squared errors (RMSE) and predictive log scores (LS). Table~\ref{tab:forecast_wholesample} reports the results for the full sample. 

Forecast accuracy varies significantly across models, variables, and forecast horizons. For real oil prices, the three models deliver broadly similar predictions. In contrast, the performance gap is more pronounced for industrial production: the FABART model consistently outperforms the linear benchmarks, achieving lower RMSEs and better log scores than both the BVAR and FAVAR across all horizons. In terms of log scores, the deterioration in forecast densities is most acute for the FAVAR at the three-month horizon, reflecting a substantial worsening in density forecasts. In contrast, FABART maintains more contained log scores, indicating a more stable forecast performance.

Turning to financial variables, the forecast performance becomes more mixed. For the excess bond premium (EBP), the BVAR achieves the lowest RMSEs at one and three months, but its log scores do not consistently outperform those of the FABART model. FAVAR performs weakest in both metrics. For the S\&P 500 index, the linear BVAR displays the strongest forecasting performance at the 1-month horizon. While results are more comparable to those of the FABART model at longer horizons, the BVAR achieves slightly better log scores throughout, particularly in the short and medium term. The FAVAR model performs notably worse across both metrics, especially in terms of density forecasting accuracy.

Taken together, the full-sample results underscore the relative strength of the FABART model, particularly in terms of forecast accuracy and density fit for macroeconomic variables. Nevertheless, the BVAR model retains some advantage in short-horizon forecasts for financial series.

\begin{table}[htp]
    \centering
    \caption{Forecast Evaluation (Whole Sample: 2014:01--2024:06)}
    \label{tab:forecast_wholesample}
    \begin{threeparttable}
        \footnotesize
        \begin{tabular}{lcccccc}
            \toprule
            & \multicolumn{2}{c}{\textbf{FABART}} & \multicolumn{2}{c}{\textbf{BVAR}} & \multicolumn{2}{c}{\textbf{FAVAR}} \\
            \cmidrule(lr){2-3} \cmidrule(lr){4-5} \cmidrule(lr){6-7}
            & RMSE & LS & RMSE & LS & RMSE & LS \\
            \midrule
            \textbf{Real Oil} & & & & & & \\
            \midrule
            1M  & 0.014 & 2.214 & 0.018 & 2.358 & 0.017 & 2.338 \\
            3M  & 0.025 & 1.458 & 0.030 & 1.616 & 0.029 & 1.630 \\
            12M & 0.051 & 1.002 & 0.061 & 0.931 & 0.058 & 1.031 \\
            \midrule
            \textbf{Industrial Production} & & & & & & \\
            \midrule
            1M  & 0.701 & -1.330 & 0.712 & -5.835 & 0.770 & -5.102 \\
            3M  & 1.192 & -1.920 & 1.234 & -5.928 & 1.323 & -10.661 \\
            12M & 1.725 & -2.502 & 2.773 & -5.872 & 3.473 & -3.897 \\
            \midrule
            \textbf{EBP} & & & & & & \\
            \midrule
            1M  & 0.233 & -0.489 & 0.155 & -0.304 & 0.314 & -1.067 \\
            3M  & 0.267 & -0.496 & 0.248 & -0.510 & 0.358 & -0.867 \\
            12M & 0.312 & -0.584 & 0.371 & -0.705 & 0.399 & -0.623 \\
            \midrule
            \textbf{S\&P 500} & & & & & & \\
            \midrule
            1M  & 77.317 & -6.542 & 77.278 & -6.137 & 62.480 & -18.276 \\
            3M  & 82.663 & -6.612 & 130.662 & -6.644 & 110.824 & -30.302 \\
            12M & 148.381 & -7.921 & 242.411 & -7.388 & 235.346 & -16.481 \\
            \bottomrule
        \end{tabular}
        \begin{tablenotes}
            \tiny
            \item \textit{Notes:} The table reports the average Root Mean Squared Errors (RMSE) and predictive log-scores (LS) for different forecasting models over the full evaluation sample (2014:01–2024:06). The last observation used for estimation is 2023:06, with forecasts evaluated against data through 2024:06.
        \end{tablenotes}
    \end{threeparttable}
\end{table}

\begin{figure}[htp]
    \centering  
    \includegraphics[width=0.75\linewidth]{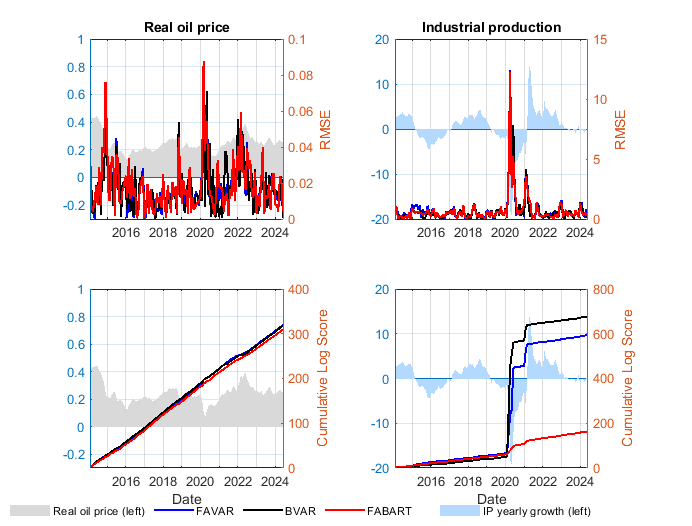}
    \caption{Forecast evaluation across time: 1-month ahead performance for macroeconomic variables}
    \label{fig:1M_TS}
\end{figure}

\begin{figure}[htp]
    \centering  
    \includegraphics[width=0.75\linewidth]{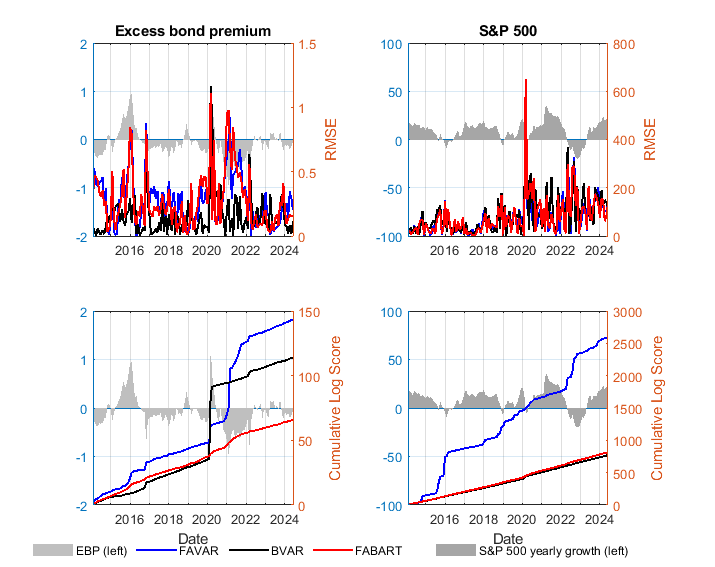}
     \caption{Forecast evaluation across time: 1-month ahead performance for financial variables}
    \label{fig:1M_TS_financial}
    \vspace{0.2em}
    \tiny{\textit{Note:} 1-month-ahead forecast evaluation results over the 2014:01–2024:06 period for the FAVAR, BVAR, and FABART models. The top row shows root mean square error (RMSE) series. The bottom row displays cumulative log scores in absolute value. Shaded areas correspond to the year-on-year growth rates for industrial production and S\&P 500, or levels for real oil prices and the excess bond premium. Lower RMSE values and cumulative log scores indicate better forecast performance.}
\end{figure}

To assess how the FABART model performed relative to linear benchmark models over time, Figures \ref{fig:1M_TS}--\ref{fig:1M_TS_financial} display the RMSE and cumulative log scores for 1-month-ahead forecasts of selected macroeconomic and financial variables over the 2014:01–2024:06 period. The top panels show the RMSEs, while the bottom panels report the cumulative log scores. Following standard practice, I compute the log score as the logarithm of the predictive density evaluated at the realized value. For comparability across models and series, I report the cumulative sum of these log scores in absolute value.\footnote{Log scores are based on the logarithm of the predictive density evaluated at the realized outcome. Reporting their cumulative sum in absolute value facilitates interpretation: lower values indicate that the model consistently assigned higher probability to outcomes that actually occurred, reflecting better density forecast performance. See  \cite{geweke2010comparing} or \cite{alessandri2017financial}.} In this convention, lower values signal superior predictive performance, as they correspond to higher likelihood assigned to the realized outcomes. Shaded areas correspond to the observed series.

The first column of Figure \ref{fig:1M_TS} presents the results for real oil prices. Forecast performance is broadly comparable across the three models—both in terms of RMSE and log score accumulation—though the FABART model (red line) performs slightly better, as indicated by a lower cumulative log score.

The second column displays the results for industrial production. Across all models, RMSEs increase sharply during the COVID-19 period and its immediate aftermath. This deterioration in forecast performance coincides with the extreme year-on-year fluctuations visible in the shaded series. As also reported in Table \ref{tab:forecast_wholesample}, average RMSE performance is broadly similar across models. However, FABART exhibits notably greater stability in density forecasts, as evidenced by a smoother and less volatile accumulation of log scores. This suggests that FABART delivers more robust predictive distributions, particularly during periods of heightened volatility.

Figure \ref{fig:1M_TS_financial} displays the forecast evaluation across time for two selected financial variables. The first column presents the results for the excess bond premium. The linear BVAR displays smaller RMSEs, but it also exhibits a notable decline in density forecast performance following the COVID period, consistent with the increased volatility in the series. By contrast, the FABART model maintains a more stable log score trajectory, underscoring its robustness in capturing predictive densities during periods of financial stress.

The second column shows the results for the S\&P 500 index. Forecast accuracy, measured by the RMSE, deteriorates sharply in early 2020 across all models. The cumulative log scores indicate that the FAVAR model performs notably worse, while the linear BVAR and FABART models exhibit similar and more stable density forecast performance.

\begin{landscape}
\begin{figure}[htp]
  \centering
  \begin{minipage}[b]{0.48\textwidth}
    \centering
    \includegraphics[width=\textwidth, height=0.65\textheight]{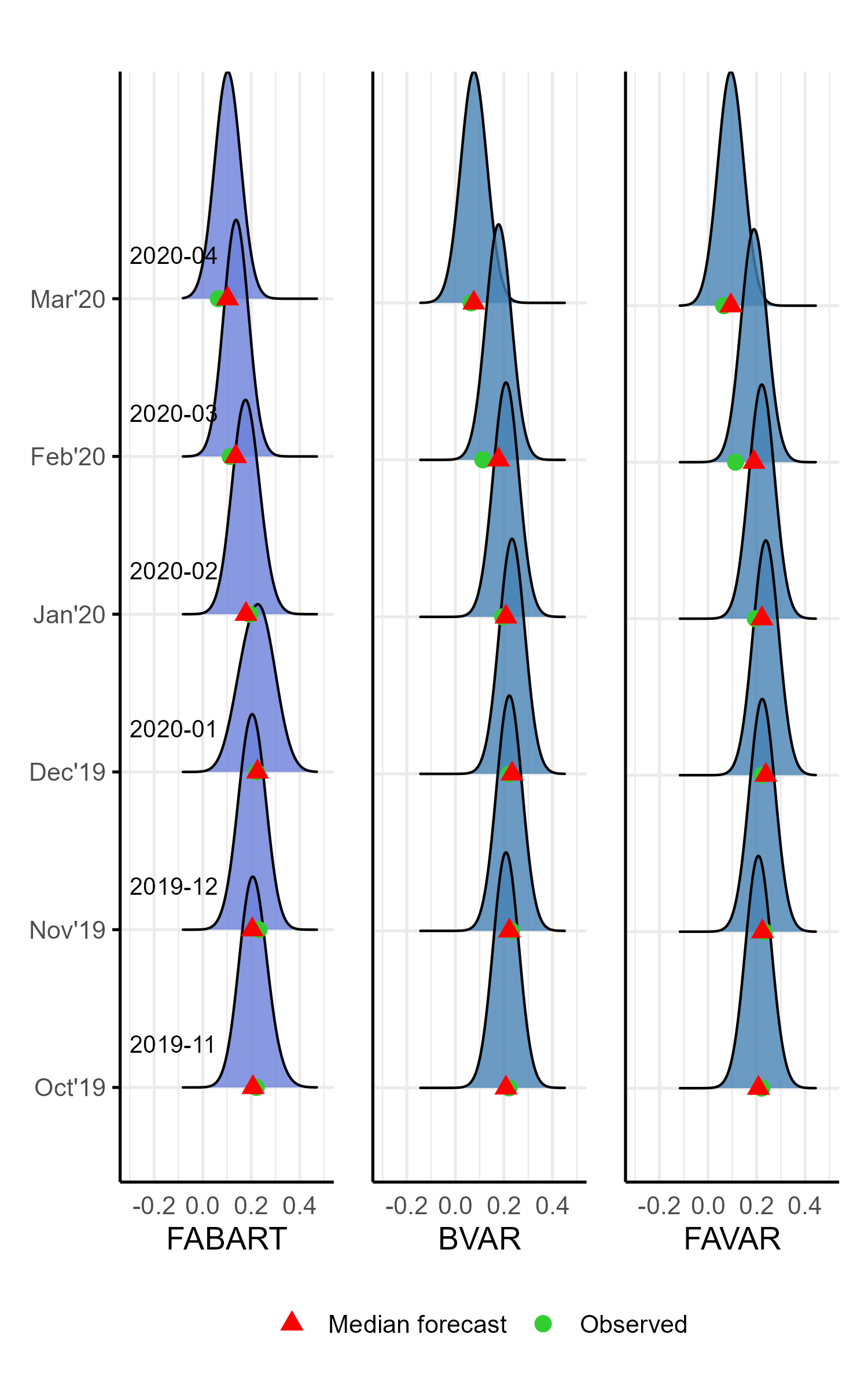}
    \caption{Real oil prices: 1-mth ahead}
    \label{fig:REOIL1M}
  \end{minipage}
  \hspace{1em} 
  \begin{minipage}[b]{0.48\textwidth}
    \centering
    \includegraphics[width=\textwidth, height=0.65\textheight]{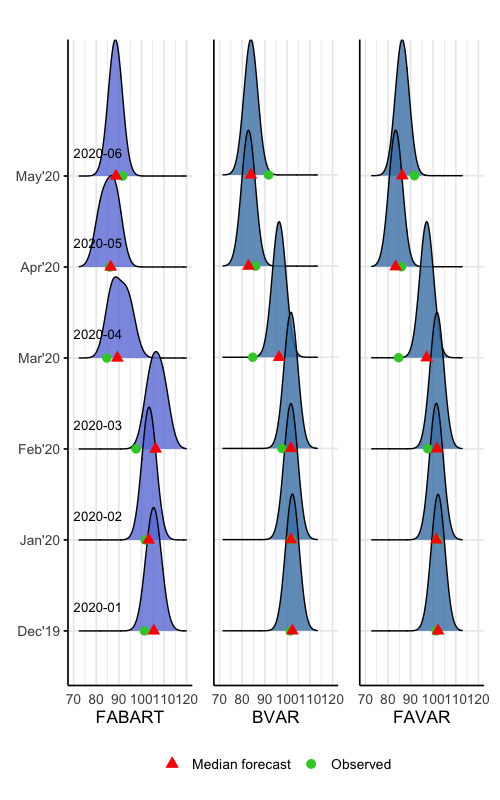}
    \caption{Industrial prod.: 1-mth ahead}
    \label{fig:INDPRO1M}
  \end{minipage}

  \vspace{1em}
  {\tiny
  \textit{Note:} Posterior densities and realized values. The y-axis indicates the last month of data included in the estimation sample. Labels above the axis refer to the forecasted period. Green dots represent the realized values, while red triangles indicate the corresponding median forecasts. LHS chart displays the forecasts for real-oil prices, RHS displays the forecasts for industrial production. Both variables are depicted in levels. 
  }
\end{figure}
\end{landscape}

This relative advantage of FABART aligns with recent literature emphasizing the limitations of linear models during extreme events. As emphasized by \cite{huber2020nowcasting}, the COVID-19 shock generated data far outside historical ranges, rendering traditional mixed-frequency VARs unreliable and motivating the adoption of nonlinear approaches. In this spirit, \cite{hauzenberger2023real} show that nonlinear dimension reduction techniques—such as autoencoders and kernel-based principal components—markedly improve forecast accuracy for inflation during turbulent periods. To examine how these modeling differences play out during episodes of heightened uncertainty, Figure \ref{fig:REOIL1M} zooms in on the critical months of Spring 2020, when both industrial production and oil markets experienced unprecedented disruptions. Industrial production saw extreme outliers, including the largest monthly change in the post-1974 period, while oil markets were hit by the combined shock of the COVID outbreak and the collapse of OPEC+ negotiations. As noted by \cite{baumeister2024risky}, traditional models failed to anticipate the severity of the oil price collapse. The figure provides a granular view of how each model’s forecast densities responded in quasi real time.

Figure \ref{fig:REOIL1M} (LHS) illustrates the one-month-ahead predictive densities for real oil prices across the different models during the onset of the COVID-19 shock. All models gradually incorporated the steep decline in oil prices, with observed values (green dots) consistently falling within the forecast bands. Over time, the posterior densities of all specifications shifted into negative territory, reflecting the sharp deterioration in oil market conditions. Among the three models, FABART exhibited a faster adjustment to the downturn, particularly between February and March 2020. Its nonlinear structure allowed for greater responsiveness to the incoming data, leading to forecast distributions that re-centered more accurately around the observed outcomes. In contrast, the linear BVAR and FAVAR models adapted more slowly, resulting in delayed recognition of the oil price collapse.

The one-month-ahead predictive densities for industrial production are presented in Figure \ref{fig:REOIL1M} (RHS). The comparative advantage of the FABART model becomes particularly evident during the early phase of the COVID-19 shock. In the forecast made in February 2020—targeting March outcomes—FABART already shows signs of a downward adjustment. This adaptability becomes even more pronounced in the March forecast (for April 2020), where the linear  models entirely fail to anticipate the sharp contraction in industrial production. In contrast, FABART captures the deterioration in real activity more accurately, with its predictive density re-centering closer to the observed outcome.

\clearpage

\subsection{Effects of oil price shocks on the U.S. economy}
\subsubsection{Aggregate effects}
A substantial literature has established that oil price increases have more pronounced macroeconomic effects than decreases, also referred to as sign asymmetry. \cite{hamilton2003oil} shows that output declines following oil price hikes are not mirrored by equivalent expansions after price declines. His results suggest that increases in oil prices—particularly those that reverse earlier declines—contain significant predictive content for GDP, whereas decreases do not. This asymmetry has been linked to mechanisms such as uncertainty, adjustment costs, and irreversible investment. Moreover, the analysis in \cite{balke2002oil} indicates that neither monetary policy nor standard price dynamics fully explain the asymmetric responses of output to oil shocks. Their findings instead highlight the role of real-side frictions—such as sectoral reallocation and capital adjustment—as key drivers of the observed nonlinearity.

To assess whether such asymmetries are present within the FABART framework, this section investigates potential nonlinearities in the transmission of oil price shocks to both the U.S. and global economy, with particular attention to the sign of the shock.\footnote{The analysis uses pre-pandemic aggregate and state-level data, as detailed in Section~\ref{statelevel}. The monthly sample spans from 1974:01 to 2016:12, aligning with recent studies on oil supply news shocks and their macroeconomic effects (e.g., \cite{kanzig2021macroeconomic,miescu2024cardiff,caravello2024disentangling,forni2023asymmetric,mumtaz2024distribution}). As noted by \cite{kanzig2021macroeconomic}, the external instrument is available only from 1984:04 onward, therefore the estimation of the $\mathbf{A}$ matrix is limited to that subsample.} The nonlinear nature of the model allows to capture the possibility that positive and negative shocks propagate through distinct channels and with differing intensity.

\begin{figure}[htp]
    \centering  
    \includegraphics[width=0.8\linewidth]{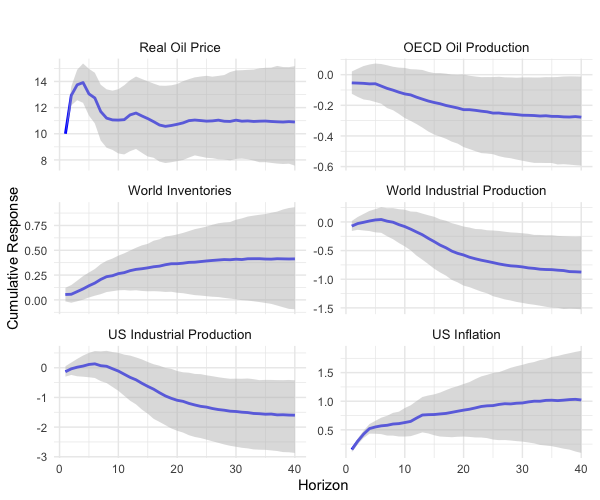} 
 
    \tiny{\textit{Note:} Generalized impulse response functions (GIRFs) of the baseline FABART model to a 10\% increase in real oil prices driven by an oil supply news shock. Shaded regions represent 16–84\% error bands.} 
  \caption{Response to a positive oil price shock.}\label{fig:BaselineGIRF}
\end{figure}

In line with \cite{kanzig2021macroeconomic}, the size of the shock is calibrated to generate a 10\% increase in the real price of oil in response to an oil supply news shock. Figure \ref{fig:BaselineGIRF} displays the median responses to this shock. Overall, the results are broadly consistent with the findings of \cite{kanzig2021macroeconomic} in terms of both the sign and magnitude of the responses. In particular, oil supply shocks lead to higher inflation and a contraction in real economic activity, consistent with standard transmission mechanisms, as also shown in \cite{mumtaz2024distribution}, who employ a FAVAR model with an external instrument to study oil price shocks. Moreover, the relevance of oil shocks for short-term inflation dynamics is supported by \cite{jacquinot2009assessment}, who, using a DSGE framework calibrated to the euro area, find that oil price changes are a key driver of short-run inflation fluctuations. Consistent with the findings of \cite{kanzig2021macroeconomic}, the shock also leads to a significant increase in global oil inventories on impact, followed by sluggish growth. This pattern reflects forward-looking behavior by market participants anticipating future scarcity or higher oil prices.

\begin{figure}[htp]
        \centering  \includegraphics[width=0.8\linewidth]{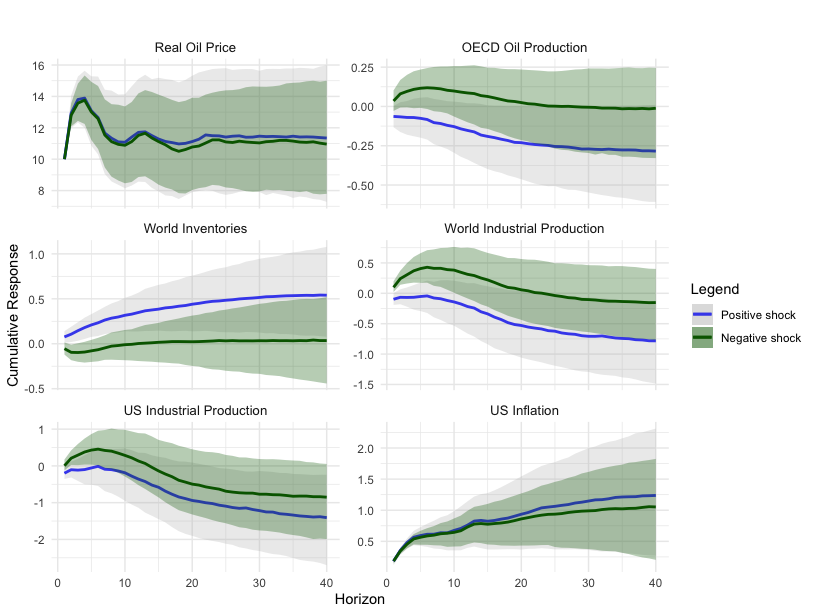}   
         
        \tiny{
        \textit{Note:} Generalized impulse response functions (GIRFs) illustrate the size asymmetry in the response of the baseline FABART model to oil price shocks. Shaded areas represent the 16–84\% credible intervals. The responses to negative shocks are mirrored to facilitate comparison with positive shocks. Blue lines show the response to a postive shock of a 10\% increase in real oil prices driven by an oil supply news shock, while green lines depict the response to a negative shock of the same size.} 
  \caption{Sign asymmetry in the response to oil price shocks.}\label{fig:AssymmGIRF}
\end{figure}

Figure \ref{fig:AssymmGIRF} displays the impulse responses to positive and negative oil price shocks, with the latter mirrored for comparability.\footnote{For clarity, the Appendix includes a separate figure showing the original responses to negative oil price shocks.} The results reveal clear sign asymmetries: positive shocks induce a persistent increase in inflation, which is notably stronger than the disinflationary response to negative shocks. However, their effects on real activity unfold more gradually—U.S. and global industrial production decline significantly only after approximately 20 months. By contrast, negative oil price shocks trigger an immediate and significant decline in industrial production, with effects lasting up to one year. This pattern challenges the conventional view that oil price reductions are expansionary. As emphasized by \cite{hamilton2003oil}, such nonlinearities may stem from asymmetries in adjustment costs, heightened uncertainty, and the irreversibility of investment in energy-intensive durable goods. While oil price increases can disrupt production plans and erode consumer confidence, declines often fail to stimulate activity symmetrically due to frictions in reallocating resources and expectations that lower prices may not persist.

In terms of other key oil market drivers considered in the analysis, a decline in oil prices leads to a modest fall in OECD oil production, significant up to the 10-month horizon, after which the response becomes centered around zero and remains insignificant until the end of the forecast horizon. In contrast, following a positive shock, oil production initially shows no significant response, but a contraction materializes after about 20 months.

Recent literature offers complementary and contrasting interpretations of such asymmetries. \cite{forni2023asymmetric}, for instance, emphasize the role of uncertainty in shaping the macroeconomic effects of oil supply news shocks. In their framework, uncertainty rises in response to both positive and negative shocks, but its macroeconomic consequences are asymmetric: it amplifies the contractionary impact of oil price increases while dampening the expansionary effects of oil price declines. This helps to explain the muted real activity responses to negative shocks in their findings. By contrast, the results presented here point to significant real effects following both positive and negative shocks. As this paper does not impose an explicit uncertainty mechanism, alternative transmission channels—such as demand reallocation, precautionary behavior, or financial frictions—may play a more dominant role in shaping the dynamics of industrial production, particularly in response to negative price shocks.

In line with this perspective, the theoretical framework developed by \cite{miescu2024cardiff} embeds uncertainty and labor market frictions into a nonlinear DSGE model with forward-looking agents and endogenous job separation. Rather than distinguishing between positive and negative shocks, their model focuses on shock size and shows that larger oil supply news shocks lead to disproportionately stronger effects on real activity, employment risk, and financial variables such as risk premia. In particular, larger shocks increase the likelihood of job loss and precautionary savings, which in turn amplify economic downturns. While the framework successfully captures inflation persistence and uncertainty-driven downturns, it does not account for the sign-dependent contractionary effects of negative shocks observed in this paper. This contrast reinforces the idea that the macroeconomic impact of oil shocks depends not only on their size or direction, but also on how risk and uncertainty are propagated through different economic frictions.

A further comparison with \cite{caravello2024disentangling} underscores the importance of distinguishing between sign and size nonlinearities. Their methodological contribution shows that local projection models augmented with even or odd functions of shocks can disentangle these two forms of asymmetry. Applying their framework to oil supply news shocks, they find evidence of size nonlinearities but not sign-dependent responses. In contrast, the results presented here suggest a clear sign asymmetry, both in the timing and magnitude of inflation and output responses. This divergence suggests that within the FABART framework, future work should further examine how shock size, uncertainty, and sign interact to drive macroeconomic asymmetries in the response to oil price movements—particularly in light of evidence by \cite{arce2025economic} showing that heightened oil supply uncertainty, even in the absence of actual supply cuts, can generate inflationary pressures and sector-specific output declines that resemble the effects typically associated with negative oil supply shocks.

\newpage

\subsubsection{State-level employment effects}
While the preceding analysis has focused on the U.S. aggregate effects of oil price shocks on a range of macro-financial variables, the following analysis shifts to the U.S. federal state level and examines employment responses.
The model underlying the state-level analysis is identical to the one previously presented, as the number of variables included in the observation equation remains unchanged.

\begin{figure}[H]
\centering
\includegraphics[width=0.85\linewidth]{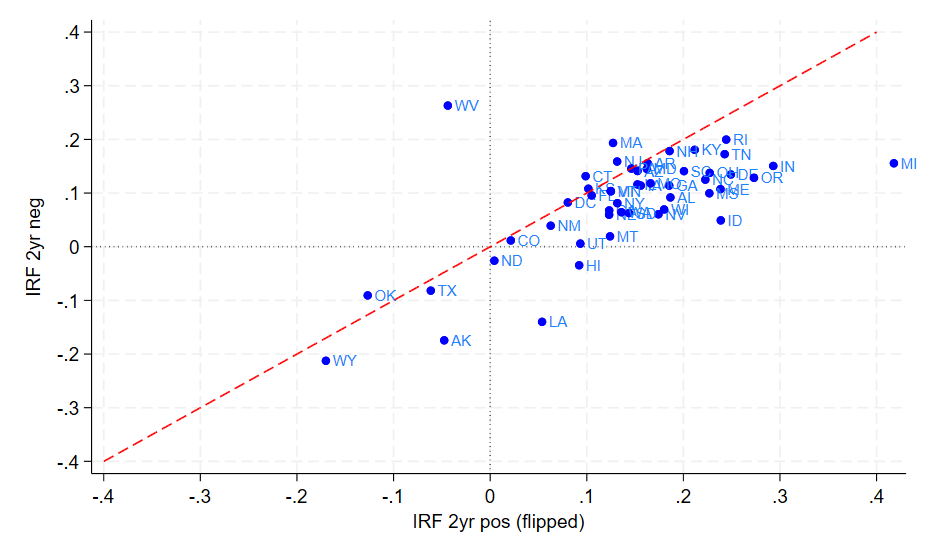}
\tiny{
\textit{Note:} This figure compares the median two-year cumulative employment contraction due to a positive oil price shock (x-axis, sign-adjusted) and a negative oil price shock (y-axis).  
Both axes are expressed in contraction space. The red dashed line represents the 45-degree benchmark for symmetric responses.  
Observations below this line indicate that employment losses following positive oil price shocks exceed the magnitude of employment responses to negative shocks.
}
\caption{Cross-state symmetry in employment responses to oil price shocks}
\label{fig:irf_flippedpos}
\end{figure}

To examine whether positive and negative shocks induce asymmetric employment contractions across U.S. states, Figure~\ref{fig:irf_flippedpos} plots the two-year cumulative employment responses. The response to positive shocks is sign-adjusted and placed on the x-axis, while the response to negative shocks is plotted on the y-axis.
The red dashed 45-degree line represents the benchmark of symmetry: states lying on this line would experience equal employment losses from both types of shocks.
The majority of states fall below the line, suggesting that employment contractions tend to be larger following positive oil price shocks than following negative ones.
This asymmetry is consistent with the findings of \cite{hamilton2003oil} and \cite{balke2002oil}, who emphasize the role of real-side frictions—such as adjustment costs, investment irreversibility, and uncertainty—in amplifying the effects of oil price increases relative to declines.

To formally assess sign asymmetry, I compare the two-year cumulative responses to positive and negative shocks across the posterior distribution.  
The results, presented in Appendix Figure~\ref{fig:Densitysign}, indicate that the posterior mass lies predominantly to the left of zero, suggesting that positive shocks induce significantly larger employment contractions than negative shocks.  
This Bayesian evidence is corroborated by a one-sample t-test on the cross-sectional distribution of state-level median cumulative responses, reported in Appendix Table~\ref{tab:ttest_irf_diff}, which rejects the null hypothesis of a zero mean difference at the 1\% significance level. 

As an additional exercise, I assess whether structural characteristics help shape the magnitude of employment responses to oil price shocks across states. Specifically, I examine whether variation in industrial composition—particularly the relative importance of manufacturing and mining sectors—can account for part of the observed cross-sectional heterogeneity. The results, presented in the Appendix (Table \ref{tab:irf_asymmetry}) , are based on a cross-sectional regression that relates the cumulative employment responses after 2 years to sectoral shares and suggest that sectoral specialization may condition the intensity of oil price shock transmission: states with larger manufacturing sectors tend to experience deeper employment contractions, while those with greater mining exposure appear more insulated. These findings are consistent with \cite{mumtaz2018state}, who document a similar role for industrial structure in shaping state-level income responses to aggregate uncertainty shocks.

Taken together, the results point to two key patterns in state-level employment responses to oil price shocks: (i)  sign asymmetries, with positive shocks triggering larger contractions than negative ones, and (ii) cross-sectional heterogeneity in the magnitude of these responses, with industrial composition playing an important role. Nevertheless, the mechanisms driving these asymmetries merit further investigation. Future research could examine how labor market frictions, regional demand conditions, or distributional channels contribute to these heterogeneous effects. Recent evidence indicates that oil shocks can have unequal consequences across income groups, suggesting an additional transmission channel via regional differences in household exposure (e.g., \cite{berisha2021income, mumtaz2024distribution}). Building on the discussion in the previous subsection, further work on the interaction between shock size, sign, and uncertainty in the transmission of oil shocks could shed light on how these dimensions jointly shape regional labor market dynamics (e.g., \cite{arce2025economic,forni2023asymmetric}).

\clearpage

\section{Conclusion}\label{sec:Conclusion}

This article develops the FABART model, a novel framework that integrates Bayesian Additive Regression Trees (BART) into a Factor-Augmented Vector Autoregressive (FAVAR) structure to forecast U.S. macro-financial variables and analyze asymmetries in the transmission of oil supply news shocks. By combining flexible nonparametric methods with latent factor structures, FABART effectively captures complex, nonlinear relationships between observables and latent factors, particularly during periods of economic and geopolitical instability.

A simulation experiment comparing FABART to linear alternatives and a Monte Carlo simulation demonstrate that the model accurately recovers the relationship between latent factors and observables in the presence of nonlinearities, while remaining consistent when the true data-generating process is linear. This robustness highlights FABART’s adaptability to different underlying structures and its ability to avoid overfitting.

The empirical application evaluates the forecasting performance of FABART relative to standard linear benchmarks across a range of U.S. macro-financial variables. Results show that FABART substantially improves forecast accuracy for industrial production, while delivering comparable performance for real oil prices and mixed results for financial variables. Importantly, FABART maintains stable forecasting performance during periods of heightened volatility, such as the COVID-19 pandemic, when the accuracy of linear models tends to deteriorate.

The article also uncovers pronounced sign asymmetries in the transmission of oil supply news shocks: positive shocks induce stronger and more persistent contractions in real activity and inflation than the expansions triggered by negative shocks. These patterns are consistent with mechanisms such as adjustment costs, investment irreversibility, and heightened uncertainty that amplify the macroeconomic effects of positive oil shocks. Extending the analysis to the U.S. state level, the results reveal similar sign asymmetries in employment responses, with positive shocks generating sharper employment contractions than the improvements following negative shocks. 

The FABART framework captures key asymmetries and nonlinear propagation channels in the transmission of oil supply news shocks. However, understanding the mechanisms driving the emergence of asymmetric responses remains an important avenue for future research. Further work could explore how adjustment costs, investment irreversibility, and heightened uncertainty amplify the contractionary effects of positive oil price shocks relative to the muted expansions following negative shocks. At the state level, a more granular analysis of sectoral specialization, regional demand dynamics, and distributional channels may help explain the observed cross-sectional heterogeneity in employment responses. Finally, extending the framework to examine the interaction between uncertainty, shock size, and sign dependence would provide additional insights into the nonlinear transmission of oil shocks across both aggregate and regional outcomes, building on the recent advances discussed in this article.


\bibliographystyle{apalike}
\bibliography{mybib}

\begin{thebibliography}{}

\bibitem[Alessandri and Mumtaz, 2017]{alessandri2017financial}
Alessandri, P. and Mumtaz, H. (2017).
\newblock {Financial conditions and density forecasts for US output and inflation}.
\newblock {\em Review of Economic Dynamics}, 24:66--78.

\bibitem[Arce-Alfaro, 2025]{arce2025economic}
Arce-Alfaro, G. (2025).
\newblock The economic implications of oil supply uncertainty.
\newblock {\em Energy Economics}, page 108425.

\bibitem[Bai and Ng, 2008]{BaiNg2008}
Bai, J. and Ng, S. (2008).
\newblock Forecasting economic time series using targeted predictors.
\newblock {\em Journal of Econometrics}, 146:304--317.

\bibitem[Balke et~al., 2002]{balke2002oil}
Balke, N.~S., Brown, S.~P., and Yucel, M.~K. (2002).
\newblock Oil price shocks and the us economy: Where does the asymmetry originate?
\newblock {\em The Energy Journal}, 23(3).

\bibitem[Ba{\'n}bura et~al., 2010]{banbura2010large}
Ba{\'n}bura, M., Giannone, D., and Reichlin, L. (2010).
\newblock Large bayesian vector auto regressions.
\newblock {\em Journal of Applied Econometrics}, 25(1):71--92.

\bibitem[Baumeister and Hamilton, 2019]{baumeister2019structural}
Baumeister, C. and Hamilton, J.~D. (2019).
\newblock Structural interpretation of vector autoregressions with incomplete identification: Revisiting the role of oil supply and demand shocks.
\newblock {\em American Economic Review}, 109(5):1873--1910.

\bibitem[Baumeister et~al., 2024]{baumeister2024risky}
Baumeister, C., Huber, F., and Marcellino, M. (2024).
\newblock Risky oil: It's all in the tails.
\newblock Technical report, National Bureau of Economic Research.

\bibitem[Baumeister and Kilian, 2015]{baumeister2015forecasting}
Baumeister, C. and Kilian, L. (2015).
\newblock Forecasting the real price of oil in a changing world: a forecast combination approach.
\newblock {\em Journal of Business \& Economic Statistics}, 33(3):338--351.

\bibitem[Baumeister et~al., 2022]{baumeister2022energy}
Baumeister, C., Korobilis, D., and Lee, T.~K. (2022).
\newblock Energy markets and global economic conditions.
\newblock {\em Review of Economics and Statistics}, 104(4):828--844.

\bibitem[Baumeister et~al., 2013]{baumeister2013changes}
Baumeister, C., Liu, P., and Mumtaz, H. (2013).
\newblock Changes in the effects of monetary policy on disaggregate price dynamics.
\newblock {\em Journal of Economic Dynamics and Control}, 37(3):543--560.

\bibitem[Berisha et~al., 2021]{berisha2021income}
Berisha, E., Chisadza, C., Clance, M., and Gupta, R. (2021).
\newblock Income inequality and oil resources: Panel evidence from the united states.
\newblock {\em Energy Policy}, 159:112603.

\bibitem[Bernanke et~al., 2005]{Bernanke2005}
Bernanke, B.~S., Boivin, J., and Eliasz, P. (2005).
\newblock Measuring the effects of monetary policy: a factor-augmented vector autoregressive (favar) approach.
\newblock {\em The Quarterly Journal of Economics}, 120(1):387--422.

\bibitem[Caravello and Martinez-Bruera, 2024]{caravello2024disentangling}
Caravello, T. and Martinez-Bruera, P. (2024).
\newblock Disentangling sign and size non-linearities.
\newblock {\em SSRN 4704050}.

\bibitem[Carter and Kohn, 1994]{carter1994gibbs}
Carter, C.~K. and Kohn, R. (1994).
\newblock On gibbs sampling for state space models.
\newblock {\em Biometrika}, 81(3):541--553.

\bibitem[Chipman et~al., 1998]{chipman1998bayesian}
Chipman, H.~A., George, E.~I., and McCulloch, R.~E. (1998).
\newblock Bayesian cart model search.
\newblock {\em Journal of the American Statistical Association}, 93(443):935--948.

\bibitem[Chipman et~al., 2010]{chipman2010bart}
Chipman, H.~A., George, E.~I., and McCulloch, R.~E. (2010).
\newblock {BART: Bayesian additive regression trees}.
\newblock {\em The Annals of Applied Statistics}, 4(1):266--298.

\bibitem[Crawford et~al., 2019]{crawford2019variable}
Crawford, L., Flaxman, S.~R., Runcie, D.~E., and West, M. (2019).
\newblock {Variable prioritization in nonlinear black box methods: A genetic association case study}.
\newblock {\em The annals of applied statistics}, 13(2):958.

\bibitem[Crawford et~al., 2018]{crawford2018bayesian}
Crawford, L., Wood, K.~C., Zhou, X., and Mukherjee, S. (2018).
\newblock Bayesian approximate kernel regression with variable selection.
\newblock {\em Journal of the American Statistical Association}, 113(524):1710--1721.

\bibitem[Del~Negro and Otrok, 2008]{del2008dynamic}
Del~Negro, M. and Otrok, C. (2008).
\newblock Dynamic factor models with time-varying parameters: measuring changes in international business cycles.
\newblock {\em FRB of New York Staff Report}, (326).

\bibitem[Doan et~al., 1984]{doan1984forecasting}
Doan, T., Litterman, R., and Sims, C. (1984).
\newblock {Forecasting and conditional projection using realistic prior distributions}.
\newblock {\em Econometric Reviews}, 3(1):1--100.

\bibitem[Dolado et~al., 2020]{dolado2020quantile}
Dolado, J.~J., Chen, L., and Gonzalo, J. (2020).
\newblock Quantile factor models.

\bibitem[Forni et~al., 2023]{forni2023asymmetric}
Forni, M., Franconi, A., Gambetti, L., and Sala, L. (2023).
\newblock Asymmetric transmission of oil supply news.
\newblock Technical report, CEPR Discussion Papers.

\bibitem[Friedman, 2001]{friedman2001greedy}
Friedman, J.~H. (2001).
\newblock Greedy function approximation: a gradient boosting machine.
\newblock {\em Annals of statistics}, pages 1189--1232.

\bibitem[Geweke and Amisano, 2010]{geweke2010comparing}
Geweke, J. and Amisano, G. (2010).
\newblock {Comparing and evaluating Bayesian predictive distributions of asset returns}.
\newblock {\em International Journal of Forecasting}, 26(2):216--230.

\bibitem[Hamilton, 2003]{hamilton2003oil}
Hamilton, J.~D. (2003).
\newblock What is an oil shock?
\newblock {\em Journal of econometrics}, 113(2):363--398.

\bibitem[Hauzenberger et~al., 2023]{hauzenberger2023real}
Hauzenberger, N., Huber, F., and Klieber, K. (2023).
\newblock Real-time inflation forecasting using non-linear dimension reduction techniques.
\newblock {\em International Journal of Forecasting}, 39(2):901--921.

\bibitem[Huber et~al., 2020]{huber2020nowcasting}
Huber, F., Koop, G., Onorante, L., Pfarrhofer, M., and Schreiner, J. (2020).
\newblock {Nowcasting in a pandemic using non-parametric mixed frequency VARs}.
\newblock {\em Journal of Econometrics}.

\bibitem[Huber and Rossini, 2022]{huber2022inference}
Huber, F. and Rossini, L. (2022).
\newblock {Inference in Bayesian additive vector autoregressive tree models}.
\newblock {\em The Annals of Applied Statistics}, 16(1):104--123.

\bibitem[Ish-Horowicz et~al., 2019]{ish2019interpreting}
Ish-Horowicz, J., Udwin, D., Flaxman, S., Filippi, S., and Crawford, L. (2019).
\newblock Interpreting deep neural networks through variable importance.
\newblock {\em arXiv preprint arXiv:1901.09839}.

\bibitem[Jacquinot et~al., 2009]{jacquinot2009assessment}
Jacquinot, P., Kuismanen, M., Mestre, R., and Spitzer, M. (2009).
\newblock An assessment of the inflationary impact of oil shocks in the euro area.
\newblock {\em The Energy Journal}, 30(1):49--84.

\bibitem[K{\"a}nzig, 2021]{kanzig2021macroeconomic}
K{\"a}nzig, D.~R. (2021).
\newblock The macroeconomic effects of oil supply news: Evidence from opec announcements.
\newblock {\em American Economic Review}, 111(4):1092--1125.

\bibitem[Kase et~al., 2025]{BIS2023}
Kase, N., Melosi, L., and Rottner, M. (2025).
\newblock Estimating nonlinear heterogeneous agent models with neural networks.
\newblock BIS Working Papers 1241, Bank for International Settlements.

\bibitem[Koop et~al., 1996]{koop1996impulse}
Koop, G., Pesaran, M.~H., and Potter, S.~M. (1996).
\newblock Impulse response analysis in nonlinear multivariate models.
\newblock {\em Journal of econometrics}, 74(1):119--147.

\bibitem[Korobilis, 2013]{korobilis2013assessing}
Korobilis, D. (2013).
\newblock Assessing the transmission of monetary policy using time-varying parameter dynamic factor models.
\newblock {\em Oxford Bulletin of Economics and Statistics}, 75(2):157--179.

\bibitem[Korobilis and Schr{\"o}der, 2024]{korobilis2024monitoring}
Korobilis, D. and Schr{\"o}der, M. (2024).
\newblock Monitoring multi-country macroeconomic risk: A quantile factor-augmented vector autoregressive (qfavar) approach.
\newblock {\em Journal of Econometrics}, page 105730.

\bibitem[Litterman, 1986a]{litterman1986forecasting}
Litterman, R.~B. (1986a).
\newblock Forecasting with bayesian vector autoregressions—five years of experience.
\newblock {\em Journal of Business \& Economic Statistics}, 4(1):25--38.

\bibitem[Litterman, 1986b]{litterman1986statistical}
Litterman, R.~B. (1986b).
\newblock A statistical approach to economic forecasting.
\newblock {\em Journal of Business \& Economic Statistics}, 4(1):1--4.

\bibitem[McCracken and Ng, 2016]{mccracken2016fred}
McCracken, M.~W. and Ng, S. (2016).
\newblock Fred-md: A monthly database for macroeconomic research.
\newblock {\em Journal of Business \& Economic Statistics}, 34(4):574--589.

\bibitem[Miescu et~al., 2024]{miescu2024cardiff}
Miescu, M., Mumtaz, H., and Theodoridis, K. (2024).
\newblock Nonlinear dynamics of large oil supply news shocks.
\newblock {\em Cardiff Economics Working Papers}.

\bibitem[Miranda-Agrippino and Ricco, 2023]{miranda2023identification}
Miranda-Agrippino, S. and Ricco, G. (2023).
\newblock Identification with external instruments in structural vars.
\newblock {\em Journal of Monetary Economics}, 135:1--19.

\bibitem[Mohammed and Mohammad, 2022]{Mohammed2022}
Mohammed, R.~I. and Mohammad, A.~A. (2022).
\newblock Stability conditions for limit cycle of smooth transition hyperbolic tangent autoregressive model.
\newblock {\em Journal of Algebraic Statistics}, 13(2):2346--2357.

\bibitem[Mumtaz, 2010]{mumtaz2010evolving}
Mumtaz, H. (2010).
\newblock {Evolving UK macroeconomic dynamics: a time-varying factor augmented VAR}.
\newblock {\em Bank of England Working Paper}.

\bibitem[Mumtaz and Piffer, 2022]{mumtaz2022impulse}
Mumtaz, H. and Piffer, M. (2022).
\newblock Impulse response estimation via flexible local projections.
\newblock {\em arXiv preprint arXiv:2204.13150}.

\bibitem[Mumtaz et~al., 2018]{mumtaz2018state}
Mumtaz, H., Sunder-Plassmann, L., and Theophilopoulou, A. (2018).
\newblock The state-level impact of uncertainty shocks.
\newblock {\em Journal of Money, Credit and Banking}, 50(8):1873--1913.

\bibitem[Mumtaz and Surico, 2012]{mumtaz2012evolving}
Mumtaz, H. and Surico, P. (2012).
\newblock Evolving international inflation dynamics: world and country-specific factors.
\newblock {\em Journal of the European Economic Association}, 10(4):716--734.

\bibitem[Mumtaz et~al., 2024]{mumtaz2024distribution}
Mumtaz, H., Theophilopoulou, A., and Drossidis, T. (2024).
\newblock The distributional effects of oil supply news shocks.
\newblock {\em Economics Letters}, 240:111769.

\bibitem[Pelger and Xiong, 2022]{PelgerXiong2022}
Pelger, M. and Xiong, Y. (2022).
\newblock State-dependent factor models for us bond yields and stock returns.
\newblock {\em Journal of Financial Economics}, 144:730--748.

\bibitem[Robertson and Tallman, 1999]{robertson1999vector}
Robertson, J.~C. and Tallman, E.~W. (1999).
\newblock {Vector autoregressions: forecasting and reality}.
\newblock {\em Economic Review-Federal Reserve Bank of Atlanta}, 84(1):4.

\bibitem[Sami and Rahman, 2021]{Sami2021}
Sami, H.~M. and Rahman, M.~M. (2021).
\newblock Determining the best activation functions for predicting stock prices in different stock exchanges through multivariable time series forecasting of lstm.
\newblock {\em American Journal of Engineering and Information Technology}, 3(1):1--12.

\bibitem[Sims, 1992]{sims1992interpreting}
Sims, C.~A. (1992).
\newblock Interpreting the macroeconomic time series facts: The effects of monetary policy.
\newblock {\em European economic review}, 36(5):975--1000.

\bibitem[Sims and Zha, 1999]{sims1999error}
Sims, C.~A. and Zha, T. (1999).
\newblock Error bands for impulse responses.
\newblock {\em Econometrica}, 67(5):1113--1155.

\bibitem[Stock and Watson, 2002]{stock2002forecasting}
Stock, J.~H. and Watson, M.~W. (2002).
\newblock Forecasting using principal components from a large number of predictors.
\newblock {\em Journal of the American statistical association}, 97(460):1167--1179.

\bibitem[Stock and Watson, 2005]{stock2005implications}
Stock, J.~H. and Watson, M.~W. (2005).
\newblock Implications of dynamic factor models for var analysis.
\newblock Technical report, National Bureau of Economic Research.

\end{thebibliography}
\clearpage

\appendix
\crefalias{section}{appendix}\label{appendix}

\section*{Priors on the VAR parameters}\label{Priors_appendix}
Following \cite{banbura2010large} a natural conjugate prior for the VAR parameters is introduced via dummy observations. The prior means $\mu_n$ are choosen as OLS estimates of the coefficients of an AR(1) regression estimated for each endogenous variable using a training sample consisting of the 40 first observations. These are  removed from the sample afterwards. The scaling factors $\sigma_n$ are set using the standard deviation of the error terms from these preliminary AR(1) regressions. Here $\iota$ reflects the degree of shrinkage which is higher the closer it is to 0. As is standard for US data, the overall prior tightness is set to $\iota = 0.1$ and a flat prior is imposed on the constant term (see, e.g., \cite{alessandri2017financial}).

\cite{litterman1986statistical} and \cite{litterman1986forecasting} proposed this priors through the application of methods of Bayesian shrinkage. I implement the Normal Inverted Wishart prior through Dummy observations as in equation (\ref{Sum_Coef}). These are set such that the moments of the Minnesota prior are matched. In that sense the prior variance decreases with increasing lag length, carrying the belief that more recent lags contain more relevant information. 
\begin{equation}\label{Sum_Coef}
Y_{D,1}
=
\begin{pmatrix}

\frac{diag(\sigma_1 \mu_1, ...,\sigma_N \mu_N )}{\iota}\\
0_{Nx(P-1)xN}
\\ ........\\
diag(\sigma_1....\sigma_N)\\ ........\\ 0_{1xN}
\end{pmatrix}, \hspace{1mm} and\hspace{2mm} X_{D,1}= \begin{pmatrix}\frac{J_P \otimes diag(\sigma_1 , ...,\sigma_N )}{\iota }  0_{NPx1}  \\  0_{NxNP} \hspace{1mm}   0_{Nx1}  \\ ........\\   0_{1xNP} \hspace{2mm}   c  \end{pmatrix} \end{equation}

Additionally, a prior on the sum of coefficients is implemented for its shown benefits in improving the models forecasting accuracy, drawing insights from \cite{sims1992interpreting}, \cite{robertson1999vector} and \cite{sims1999error}. This is a modification of the Minnesota prior suggested by \cite{doan1984forecasting} and carries the belief, that the sum of the coefficients of the lags equates to 1 (\cite{robertson1999vector}). The tightness of the sum of coefficients prior is set as in \cite{banbura2010large} $\lambda =10\iota$ and is introduced by adding the following dummy observations:

\begin{equation*}
X_{D,2}= 
\begin{pmatrix} 
\frac{(1\hspace{2mm} 2 ...p) \otimes diag(\sigma_1 \mu_1, ...,\sigma_N \mu_N )}{\lambda } & 0_{nx1}
\end{pmatrix}
\end{equation*}

\section*{Supplementary figures: Simulation exercise}
This subsection displays the simulated data underlying the results reported in Section~\ref{sec:Data}.

   \begin{figure*}[htp]
    \centering
        \includegraphics[width=0.8\textwidth]{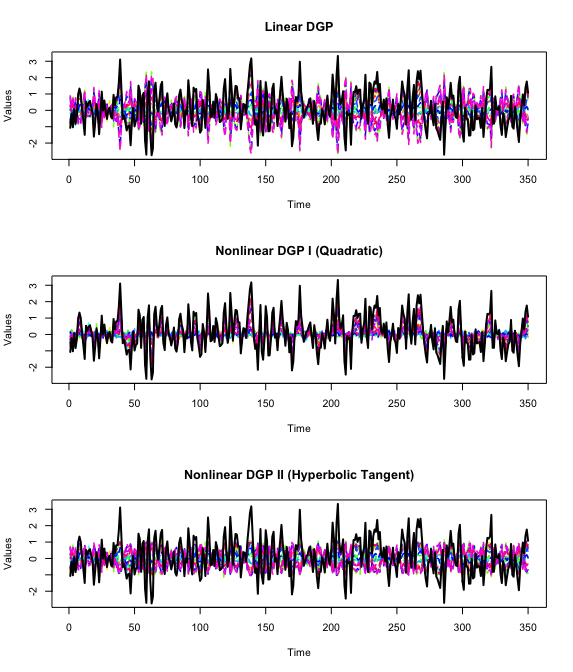} 
        \caption{Linear and non-linear DGPs based on one unobservable Factor. Straight black line depicts the unobserved Factor, colored lines the observed $X_t$ series under the different transformations.}
        \end{figure*}

\clearpage

\section*{Monte Carlo exercise}
This subsection displays the results of a Monte Carlo experiment based on 100 replications of each data-generating process (DGP)—in which the latent factor is held fixed while the observable variables \(X_t\) vary across replications—provides a deeper assessment of the properties of the FABART model. In that sense it expands on the simulation exercise in section \ref{sec:Data} as it keeps the equation fixed while allowing the observable variables to vary for each replication.

As in section \ref{sec:Data} the factor evolves according to an autoregressive process:
\begin{equation}
    F_t = c + b_1 F_{t-1}+ b_2 F_{t-2}+ b_3 F_{t-3} + e_t, \quad e_t \sim \mathcal{N}(0, \sigma^2)
\end{equation}
where \( e_t \) is normally distributed with variance \(\sigma^2 \). The coefficient vector \( \beta \) is parameterized as \( \beta = (0.6, -0.3, 0.2) \), implying that the latent factor follows an autoregressive process of order three. 

The setup follows a standard factor model in which the $N$=20 observed variables \( X_t \) is driven by $J$=1 latent factor \( F^{J}_t \). The baseline (linear) specification is given by:
\begin{equation}\label{MCeq1}
    X_t = B F^{J}_t + V_t, \quad V_t \sim \mathcal{N}(0, R),
\end{equation}

where  \( X_t \) is a \( 20 \times T \) vector of observed variables, \( B \) is a \(20 \times 1 \) factor loading matrix, and \( V_t \) follows a normal distribution with mean zero and diagonal covariance matrix \( R \) (\( 20 \times 20 \)).

The setup follows a standard factor model in which the $N=20$ observed variables \(X_t\) are driven by $J=1$ latent factor \(F^{J}_t\). The baseline (linear) specification is given by:
\begin{equation}\label{MCeq1}
X_t = B F^{J}_t + V_t, \quad V_t \sim \mathcal{N}(0, R),
\end{equation}
where \(X_t\) is a \(20 \times T\) vector of observed variables, \(B\) is a \(20 \times 1\) factor loading matrix, and \(V_t\) follows a normal distribution with mean zero and diagonal covariance matrix \(R\) (\(20 \times 20\)). The elements of \(B\) are drawn independently from a uniform distribution over the interval \([-0.9, 0.9]\). The diagonal elements of \(R\), representing the idiosyncratic variances of each series, are independently sampled from a uniform distribution. Both \(B\) and \(R\) are redrawn for each replication.

As outlined in section \ref{sec:Data} to introduce nonlinearities, Equation \ref{MCeq1} is extended in two ways:  
\begin{itemize}  
    \item[(i)] \textbf{Quadratic nonlinearity in the factor loadings:}  
    The factor loading is squared to capture asymmetric amplification effects:  
    \begin{equation}  
        X_t = B^2 F_t + V_t.  
    \end{equation}  

    \item[(ii)] \textbf{Hyperbolic tangent transformation for bounded nonlinear effects:}  
    A hyperbolic tangent transformation is applied to the factor loading to allow for smooth saturation at extreme values, mimicking nonlinear adjustment dynamics observed in financial and macroeconomic series:\footnote{The use of \(\tanh\) in time series modeling has been explored in nonlinear autoregressive models \cite{Mohammed2022} and machine learning-based forecasting \cite{Sami2021}. Additionally, \cite{BIS2023} demonstrate the effectiveness of neural network-based approaches, which frequently employ the \(\tanh\) function, in estimating nonlinear heterogeneous agent models.}  
    \begin{equation}  
        X_t = \tanh(B F_t) + V_t.  
    \end{equation}  
\end{itemize}

\begin{figure}[htp]
    \center
    \includegraphics[width=1\textwidth]{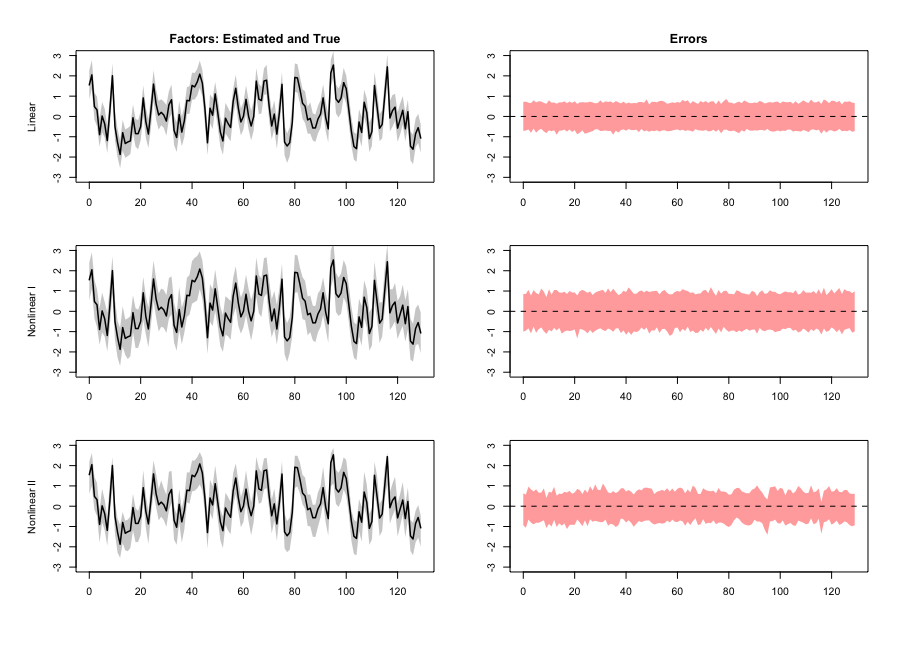}  
    \tiny{
  \textit{Note:}Posterior densities for the latent factor $F_t$ are obtained using the FABART methodology. The results are based on 100 replications for each data-generating process. Forecast errors are calculated as the difference between the estimated and true factor values for each posterior draw. All forecasts have been standardized to ensure comparability across replications and models.}
 \caption{Comparison of Posterior Factor Estimates with the True Factor}\label{fig:distributionFE}
\end{figure}

Figure \ref{fig:distributionFE} presents the distribution of estimation errors for the latent factor \(F_t\), obtained using the FABART methodology. The results are based on 100 replications for each data-generating process (DGP). The first column illustrates the posterior estimates of \(F_t\) across replications and the true factor, while the second column displays the corresponding estimation errors, defined as the difference between each posterior draw and the true factor. The figure shows that, on average, the errors are centered around zero across all DGPs and time periods. The linear DGP yields smaller estimation errors, while the magnitude of the errors increases slightly under nonlinear DGPs. These findings indicate that the MCMC algorithm accurately recovers the latent factor, demonstrating its robustness across different data-generating environments.

\clearpage

\section*{Empirical Application}

\subsection*{Fred-MD dataset}
 The dataset adheres to the format outlined in \cite{stock2002forecasting} concerning the series number, series mnemonic, data span, and transformation. Transformation codes utilized are as follows: 1 – no transformation; 2 – first difference; 4 – logarithm; 5 – first difference of logarithm. Second differencing of logarithms was not utilized. These series were directly obtained from the Fred-MD website. All variables cover the sample span of $1974:M01-2024:M08$.
\begin{table}[H]
\caption{FRED-MD Dataset Variables}\label{fred}
\centering
\small
\begin{tabular}{@{}llll@{}}
\multicolumn{4}{l}{\textbf{Group 1: Real Output and Income}} \\
1 & RPI & 5 & Real Personal Income \\
2 & W875RX1 & 5 & Real Personal Income Excluding Transfers \\
3 & INDPRO & 5 & Industrial Production Index \\
4 & IPFPNSS & 5 & IP: Final Products and Nonindustrial Supplies \\
5 & IPFINAL & 5 & IP: Final Products (Market Group) \\
6 & IPCONGD & 5 & IP: Consumer Goods \\
7 & IPDCONGD & 5 & IP: Durable Consumer Goods \\
8 & IPNCONGD & 5 & IP: Nondurable Consumer Goods \\
9 & IPBUSEQ & 5 & IP: Business Equipment \\
10 & IPMAT & 5 & IP: Materials \\
11 & IPDMAT & 5 & IP: Durable Materials \\
12 & IPNMAT & 5 & IP: Nondurable Materials \\
13 & IPMANSICS & 5 & IP: Manufacturing (SIC) \\
14 & IPB51222s & 5 & IP: Residential Utilities \\
15 & IPFUELS & 5 & IP: Fuels \\
16 & NAPMPI & 1 & ISM Manufacturing: Production Index \\
17 & CUMFNS & 2 & Capacity Utilization: Manufacturing \\

\multicolumn{4}{l}{\textbf{Group 2: Labor Market}} \\
18 & CLF16OV & 5 & Civilian Labor Force \\
19 & CE16OV & 5 & Civilian Employment \\
20 & UNRATE & 2 & Civilian Unemployment Rate \\
21 & UEMPMEAN & 2 & Average Duration of Unemployment \\
22 & UEMPLT5 & 5 & Unemployed Less Than 5 Weeks \\
23 & UEMP5TO14 & 5 & Unemployed for 5–14 Weeks \\
24 & UEMP15OV & 5 & Unemployed 15+ Weeks \\
25 & UEMP15T26 & 5 & Unemployed 15–26 Weeks \\
26 & UEMP27OV & 5 & Unemployed 27+ Weeks \\
27 & PAYEMS & 5 & Total Nonfarm Payroll Employment \\
28 & USGOOD & 5 & Goods-Producing Employment \\
29 & CES1021000001 & 5 & Mining Employment \\
30 & USCONS & 5 & Construction Employment \\
31 & MANEMP & 5 & Manufacturing Employment \\
32 & DMANEMP & 5 & Durable Goods Manufacturing Employment \\
33 & NDMANEMP & 5 & Nondurable Goods Manufacturing Employment \\
34 & SRVPRD & 5 & Service-Providing Employment \\
35 & USTPU & 5 & Trade, Transportation, and Utilities Employment \\
36 & USWTRADE & 5 & Wholesale Trade Employment \\
\end{tabular}
\end{table}

\begin{table}[H]
\centering
\small
\begin{tabular}{@{}llll@{}}
37 & USTRADE & 5 & Retail Trade Employment \\
38 & USFIRE & 5 & Financial Activities Employment \\
39 & USGOVT & 5 & Government Employment \\
40 & CES0600000007 & 1 & Avg Weekly Hours: Goods-Producing \\
41 & AWOTMAN & 2 & Avg Weekly Overtime Hours: Manufacturing \\
42 & AWHMAN & 1 & Avg Weekly Hours: Manufacturing \\
43 & NAPMEI & 1 & ISM Manufacturing: Employment Index \\
44 & CES0600000008 & 5 & Avg Hourly Earnings: Goods-Producing \\
45 & CES2000000008 & 5 & Avg Hourly Earnings: Construction \\
46 & CES3000000008 & 5 & Avg Hourly Earnings: Manufacturing \\

\multicolumn{4}{l}{\textbf{Group 3: Housing Starts and Permits}} \\
47 & HOUST & 4 & Housing Starts: Total \\
48 & HOUSTNE & 4 & Housing Starts: Northeast \\
49 & HOUSTMW & 4 & Housing Starts: Midwest \\
50 & HOUSTS & 4 & Housing Starts: South \\
51 & HOUSTW & 4 & Housing Starts: West \\
52 & PERMIT & 4 & Building Permits: Total \\
53 & PERMITNE & 4 & Building Permits: Northeast \\
54 & PERMITMW & 4 & Building Permits: Midwest \\
55 & PERMITS & 4 & Building Permits: South \\
56 & PERMITW & 4 & Building Permits: West \\

\multicolumn{4}{l}{\textbf{Group 4: Consumption, Orders, and Inventories}} \\
57 & DPCERA3M086SBEA & 5 & Real Personal Consumption Expenditures \\
58 & ACOGNO & 5 & New Orders: Consumer Goods \\
59 & NAPM & 1 & ISM Manufacturing: Composite Index \\
60 & NAPMNOI & 1 & ISM Manufacturing: New Orders Index \\
61 & NAPMSDI & 1 & ISM Manufacturing: Supplier Deliveries Index \\
62 & NAPMII & 1 & ISM Manufacturing: Inventories Index \\
\end{tabular}
\end{table}

\begin{table}[H]
\centering
\small
\begin{tabular}{@{}llll@{}}
\multicolumn{4}{l}{\textbf{Group 5: Money and Credit}} \\
63 & M1SL & 5 & Money Stock: M1 \\
64 & M2SL & 5 & Money Stock: M2 \\
65 & M2REAL & 5 & Real M2 Money Stock \\
66 & BOGMBASE & 5 & Monetary Base \\
67 & TOTRESNS & 5 & Total Reserves \\
68 & NONBORRES & 7 & Nonborrowed Reserves \\
69 & BUSLOANS & 5 & C\&I Loans at Commercial Banks \\
70 & REALLN & 5 & Real Estate Loans at Commercial Banks \\
71 & NONREVSL & 5 & Nonrevolving Consumer Credit \\
72 & MZMSL & 5 & Money Stock: MZM \\
73 & DTCOLNVHFNM & 5 & Motor Vehicle Loans Outstanding \\
74 & DTCTHFNM & 5 & Consumer Loans and Leases \\
75 & INVEST & 5 & Securities in Bank Credit \\

\multicolumn{4}{l}{\textbf{Group 6: Interest Rates and Exchange Rates}} \\
76 & FEDFUNDS & 2 & Effective Federal Funds Rate \\
77 & TB3MS & 2 & 3-Month Treasury Bill \\
78 & TB6MS & 2 & 6-Month Treasury Bill \\
79 & GS1 & 2 & 1-Year Treasury Rate \\
80 & GS5 & 2 & 5-Year Treasury Rate \\
81 & GS10 & 2 & 10-Year Treasury Rate \\
82 & AAA & 2 & Moody's Aaa Corporate Bond Yield \\
83 & BAA & 2 & Moody's Baa Corporate Bond Yield \\
84 & TB3SMFFM & 1 & 3M Treasury Minus Fed Funds \\
85 & TB6SMFFM & 1 & 6M Treasury Minus Fed Funds \\
86 & T1YFFM & 1 & 1Y Treasury Minus Fed Funds \\
87 & T5YFFM & 1 & 5Y Treasury Minus Fed Funds \\
88 & T10YFFM & 1 & 10Y Treasury Minus Fed Funds \\
89 & AAAFFM & 1 & Aaa Corporate Minus Fed Funds \\
90 & BAAFFM & 1 & Baa Corporate Minus Fed Funds \\
91 & TWEXAFEGSMTH & 5 & Trade-Weighted Dollar Index \\
\end{tabular}
\end{table}

\begin{table}[H]
\centering
\small
\begin{tabular}{@{}llll@{}}
\multicolumn{4}{l}{\textbf{Group 7: Prices}} \\
92 & WPSFD49207 & 5 & PPI: Finished Goods \\
93 & WPSFD49502 & 5 & PPI: Finished Consumer Goods \\
94 & WPSID61 & 5 & PPI: Intermediate Materials \\
95 & WPSID62 & 5 & PPI: Crude Materials \\
96 & PPICMM & 5 & PPI: Metals and Metal Products \\
97 & NAPMPRI & 1 & ISM Manufacturing: Prices Index \\
98 & CPIAUCSL & 5 & CPI: All Items \\
99 & CPIAPPSL & 5 & CPI: Apparel \\
100 & CPITRNSL & 5 & CPI: Transportation \\
101 & CPIMEDSL & 5 & CPI: Medical Care \\
102 & CUSR0000SAC & 5 & CPI: Commodities \\
103 & CUUR0000SAD & 5 & CPI: Durables \\
104 & CUSR0000SAS & 5 & CPI: Services \\
105 & CPIULFSL & 5 & CPI: All Items Less Food \\
106 & CUUR0000SA0L2 & 5 & CPI: All Items Less Shelter \\
107 & CUSR0000SA0L5 & 5 & CPI: All Items Less Medical Care \\
108 & PCEPI & 5 & PCE Price Index \\
109 & DDURRG3M086SBEA & 5 & PCE: Durable Goods \\
110 & DNDGRG3M086SBEA & 5 & PCE: Nondurable Goods \\
111 & DSERRG3M086SBEA & 5 & PCE: Services \\

\multicolumn{4}{l}{\textbf{Group 8: Stock Market}} \\
112 & S\&P 500 & 5 & S\&P Composite Price Index \\
113 & S\&P div yield & 2 & S\&P Dividend Yield \\
114 & S\&P PE ratio & 5 & S\&P Price-Earnings Ratio \\
115 & VIXCLS & 1 & CBOE Volatility Index (VIX) \\
116 & EBP & 1 & Excess Bond Premium  \\
\end{tabular}
\end{table}

\subsection*{Data for the linear BVAR model}
Table \ref{tab:redvariables} provides an overview of the variables included in the linear BVAR model, which is estimated to assess its forecasting performance relative to the proposed FABART model and the linear FAVAR approach.

In the selection of data for this analysis, I follow \cite{baumeister2015forecasting}, \cite{kanzig2021macroeconomic}, \cite{miescu2024cardiff} and \cite{baumeister2024risky}. 
The partial dependence analysis by \cite{baumeister2024risky} highlights key nonlinear relationships influencing oil price predictions. While moderate changes in predictors like global fuel consumption and petroleum inventories have limited effects, extreme variations, such as sharp contractions in the GECON indicator, steep declines in rig counts, or significant rises in the gasoline-Brent spread, induce disproportionate and asymmetric impacts on oil prices.

All variables are measured at a monthly frequency and are transformed to growth rates by taking the first difference of their natural logarithms. Exceptions to this transformation are the GECON indicator, which is inherently stationary due to its construction, and the real oil price, which is included in levels for the forecasting exercise, consistent with the approach in \cite{baumeister2024risky}.

\begin{table}[htb!]
\centering
\caption{Reduced set of variables (1974M1--2024M08)}
\label{tab:redvariables}
\begin{threeparttable}
\begin{tabularx}{\textwidth}{lXl}
\hline
\textbf{Variable} & \textbf{Description} & \textbf{Source} \\ 
\hline
Real Oil Price & WTI spot price, deflated by U.S. CPI. & EIA\tnote{a}, FRED\tnote{b} \\ 
Global Oil Production & Total global oil production. & IES\tnote{c} \\ 
OECD Petroleum Inventories & Crude oil and petroleum product inventories in OECD countries. & EIA\tnote{a}, IES\tnote{c} \\ 
Global Industrial Production & Monthly global industrial output. \small{\cite{baumeister2019structural}} &  webpage\tnote{d} \\ 
U.S. Industrial Production & Industrial output index in the U.S. & FRED\tnote{b} \\ 
U.S. CPI & Consumer Price Index. & FRED\tnote{b} \\ 
GECON Indicator & Global economic conditions indicator, reflecting economic, financial, and geopolitical factors. \small{\cite{baumeister2022energy}} & webpage\tnote{d} \\ 
\hline
\end{tabularx}
\begin{tablenotes}
\footnotesize
\item[a] Energy Information Administration (EIA).
\item[b] Federal Reserve Economic Data (FRED).
\item[c] International Energy Statistics (IES).
\item[d] Christiane Baumeister's webpage (\url{https://sites.google.com/site/cjsbaumeister/datasets}).
\end{tablenotes}
\end{threeparttable}
\end{table}



\clearpage
\subsection*{Supplementary results: Empirical analysis}
This subsection provides additional results from the empirical application.\\

\noindent\textbf{Aggregate responses}
\vspace{1ex}

\begin{figure*}[htp!]
    \centering
    \begin{minipage}{0.8\textwidth}
        \centering  
        \includegraphics[width=\linewidth]{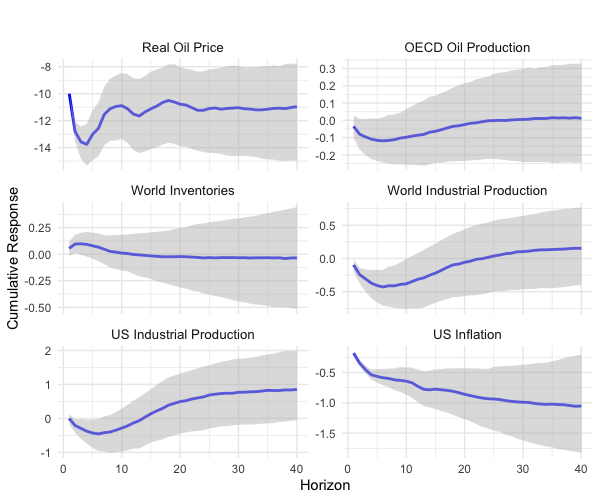} 
    \end{minipage}
    
 \tiny{
  \textit{Note:} Generalized impulse response functions (GIRFs)to a negative shock of a 10\% decrease in real oil prices driven by an oil supply news shock. 
  Shaded areas represent the 16–84\% credible intervals. }
   \caption{Responses to a negative oil price shock. }
    \label{GIRFNEG}
\end{figure*}

\noindent\textbf{State-level analysis}
\vspace{1ex}

\begin{figure}[H]
\centering
\includegraphics[width=0.85\linewidth]{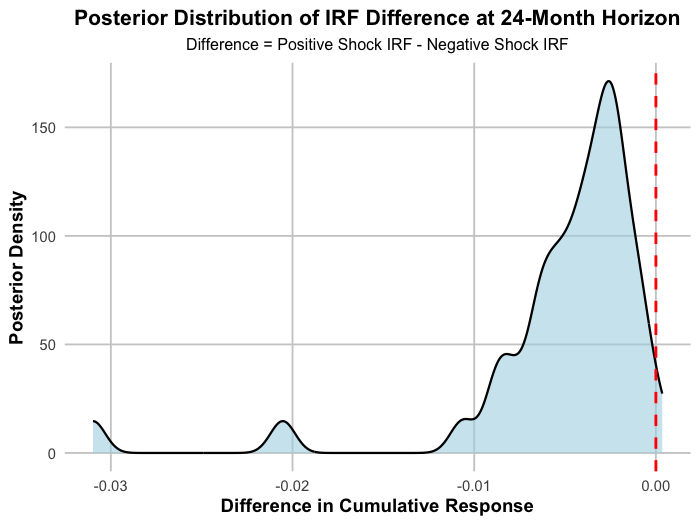}

\tiny{\textit{Note:} The figure shows the posterior density of the average cumulative response difference between positive and negative oil price shocks.  
For each posterior draw and each state, the cumulative two-year response difference is computed and subsequently averaged across states.  
The dashed vertical line at zero indicates the point of symmetry.  
A posterior distribution centered to the left of zero suggests that positive shocks induce larger employment contractions than negative shocks.}
\caption{Posterior Density of State-Level Employment Response Differences to Oil Price Shocks}
\label{fig:Densitysign}
\end{figure}

To formally assess the presence of sign asymmetry in the magnitude of employment responses, I compare the two-year cumulative response to a positive oil price shock with the sign-adjusted response to a negative shock, ensuring that both shocks are expressed as contractions.  
Figure \ref{fig:Densitysign} displays the posterior density of the average cumulative difference across states.  
The posterior mass lies predominantly to the left of zero, suggesting that positive shocks induce significantly larger employment contractions than negative shocks.  
This asymmetry is confirmed by a one-sample t-test on the state-level median cumulative responses, reported in Table \ref{tab:ttest_irf_diff}, which rejects the null hypothesis of a zero mean difference at the 1\% level.

\begin{table}[H]
\centering
\caption{Test of Mean Difference in IRFs }
\label{tab:ttest_irf_diff}
\begin{tabular}{lccc}
\toprule
 &  Mean Difference & Standard error & p-value\\
\midrule
IRF$_{\text{pos}}$ -- IRF$_{\text{neg}}$  & $-0.055$ & 0.012 & $< 0.001$ \\
\bottomrule
\end{tabular}

\tiny{\textit{Note:} This table reports the results of a one-sample t-test on the mean difference between the two-year cumulative employment response to a positive oil price shock and the sign-flipped response to a negative shock.  A negative value indicates that positive oil shocks generate larger contractions than negative shocks.}
\end{table}

Beyond sign asymmetries, I examine how differences in industrial composition across states shape the magnitude of employment responses to oil price shocks. Following the empirical strategy of \cite{mumtaz2018state}, I assess whether sectoral specialization—particularly exposure to manufacturing and mining sectors—predicts the strength of cumulative employment contractions. 

Table~\ref{tab:irf_asymmetry} presents regression results relating the two-year cumulative employment responses to state-level shares of manufacturing and mining.  
The results reveal a consistent pattern across shock types: states with a larger manufacturing sector experience more severe employment contractions, while states with greater mining activity are relatively insulated.

Specifically, a one-percentage-point increase in the manufacturing share is associated with a 0.57 percentage point larger employment contraction following a positive shock and a 0.44 percentage point contraction following a negative shock.  
In contrast, mining shares are positively associated with employment outcomes, indicating that resource-intensive states benefit from oil price increases and are less vulnerable to declines, with similar magnitudes across shocks. A similar cross-sectional pattern is found by \cite{mumtaz2018state}, who show that states with a higher manufacturing share suffered larger declines in real income following aggregate uncertainty shocks, while those with greater mining activity were more insulated.\footnote{The magnitudes and signs of the coefficients estimated in this paper appear broadly comparable to those found by \cite{mumtaz2018state}, who study the effects of aggregate uncertainty shocks across U.S.\ states. They show that states with larger manufacturing shares experience greater two-year declines in both real income and employment, while those with greater mining shares are more insulated. In cross-sectional regressions, a one-percentage-point increase in manufacturing share is associated with a 0.38--0.45 percentage point larger employment decline, and a one-point increase in mining share with a 0.47--0.55 percentage point smaller decline.}

\begin{table}[H]
\centering
\begin{threeparttable}
\caption{Cross-Sectional Regression Results}
\label{tab:irf_asymmetry}
\begin{tabular}{lcc}
\toprule
 & (1) IRF 2yr + & (2) IRF 2yr $-$ \\
\midrule
Manufacturing & $-$0.569$^{***}$ & $-$0.441$^{***}$ \\
              & (0.140)          & (0.121) \\
Mining        & \phantom{$-$}0.919$^{***}$ & \phantom{$-$}0.884$^{***}$ \\
              & (0.152)          & (0.200) \\
\midrule
Observations  & 50               & 50 \\
Adjusted $R^2$ & 0.642           & 0.629 \\
\bottomrule
\end{tabular}

\begin{tablenotes}
\tiny
\item \textit{Notes:} Each column reports OLS estimates from regressions of the two-year cumulative employment response to a positive (column 1) or negative (column 2) oil price shock on state-level manufacturing and mining shares.  
In column 2, the sign of the negative oil shock responses has been flipped so that both columns are expressed as employment contractions.  
Robust standard errors are in parentheses. $^{*}p<0.10$, $^{**}p<0.05$, $^{***}p<0.01$.
\end{tablenotes}
\end{threeparttable}
\end{table}


\end{document}